\documentclass[11pt]{article}
\usepackage{latexsym,amsthm,amssymb,amsfonts,amsbsy,graphicx,lscape,array,citesort,multibox}
\usepackage{calrsfs}

\def\a{\alpha}
\def\b{\beta}
\def\g{\gamma}
\def\c{\gamma}
\def\d{\delta}
\def\h{\eta}
\def\l{\lambda}

\def\m{\mu}
\def\n{\nu}
\def\r{\rho}
\def\o{\omega}
\def\s{\sigma}

\def\t{\tau}

\def\ve{\varepsilon}

\def\e{\varepsilon}
\def\pa{\partial}
\def\6{\partial}

\def\be{\begin{equation}}
\def\ee{\end{equation}}
\def\beq{\begin{eqnarray}}
\def\eeq{\end{eqnarray}}

\def\cc{{\cal C}}

\def\ck{{\cal K}}
\def\cl{{\cal L}}

\def\co{{\cal O}}

\def\cs{{\cal S}}

\newcommand{\bqn}{\begin{eqnarray}}\newcommand{\eqn}{\end{eqnarray}}

\newtheorem{prop}{Proposition}

\usepackage{color}
\definecolor{rougef}{rgb}{0.56,0,0}
\definecolor{vertf}{rgb}{0,0.5,0}
\definecolor{bleuf}{rgb}{0,0,0.8}

\newcommand{\TBC}[1]{\textcolor{rougef}{\textbf{[TO BE CHECKED!!]}}}
\begin{document} 
\vspace{.6cm}
\vspace*{2cm}

\begin{centering}

{\Large {\bf Consistent couplings between spin-2 and spin-3 massless fields}}

\vspace{1cm}

\begin{center}
{Nicolas Boulanger\footnote{Charg\'e de Recherches FNRS (Belgium); \tt{nicolas.boulanger@umh.ac.be}}
and Serge Leclercq\footnote{\tt{serge.leclercq@umh.ac.be}
} }
\end{center}

{\small{
\begin{center}
Universit\'e de Mons-Hainaut, Acad\'emie Wallonie-Bruxelles, \\
Service de M\'ecanique et Gravitation,\\ 
6 avenue du Champ de Mars, 7000 Mons (Belgium)\\
\end{center}}}
\end{centering}

\vspace*{2cm}

\begin{abstract}
We solve the problem of constructing consistent first-order cross-interactions between
spin-$2$ and spin-$3$ massless fields in flat spacetime of 
arbitrary dimension $n > 3$ and in such a way that the deformed gauge
algebra is non-Abelian. No assumptions are made on the number of derivatives
involved in the Lagrangian, except that it should be finite. 
Together with locality, we also impose manifest Poincar\'e invariance, 
parity invariance and analyticity of the deformations in the coupling constants. 
\end{abstract}

\newpage

%\maketitle  IS IGNORED %%%%%%%%%%%

\section{Introduction}
\label{sec:Introduction}

Although free higher-spin ($s>2$) gauge field theories are by now fairly well understood, the
Fronsdal programme \cite{Fronsdal:1978rb} which consists in introducing (non-Abelian)  
consistent interactions among these fields at the level of the action is still not achieved. 
Consistent nonlinear field equations for massless totally symmetric higher-spin fields 
in $AdS_n$ background have been constructed \cite{Vasiliev:2004qz}, which represents
a considerable achievement in higher-spin gauge field theory. 
Nonetheless, a corresponding action principle is lacking. 

In this paper we adopt the metric-like formulation for higher-spin gauge fields \cite{Fronsdal:1978rb,deWit:1979pe}, consider collections of totally symmetric rank-2 and rank-3 gauge fields in flat space of arbitrary dimension $n>3$ and study the problem of introducing non-Abelian consistent cross-interactions among spin-2 and spin-3 massless fields. 
By ``non-Abelian'', we mean that we focus on consistent deformations of the free theory 
such that the deformed gauge algebra becomes non-Abelian. 

Demanding Poincar\'e invariance and locality, the non-Abelian self-interacting problems for collections of massless spin-2 and spin-3 fields, were respectively  
investigated in \cite{Boulanger:2000rq,Boulanger:2000ni} and 
\cite{Bekaert:2005jf,Boulanger:2005br} by using the exhaustive BRST-BV cohomological 
method developed in \cite{Barnich:1993vg,Henneaux:1997bm}. 
The works \cite{Boulanger:2000rq,Boulanger:2000ni,Bekaert:2005jf,Boulanger:2005br} 
revealed the existence of manifestly covariant cubic vertices which had not previously been written before. We refer to these works and \cite{Boulanger:2001vr,Bekaert:2006us}
for details and reviews on the self-interacting non-Abelian problem for spin-2 and 
spin-3 gauge fields in flat $n$-dimensional spacetime. 
The search for consistent higher-spin cubic vertices is a very important problem and other 
approaches exist. See e.g. \cite{Metsaev:2005ar} for a recent light-cone analysis in flat spacetime and more references on the problem of consistent higher-spin vertices, including 
Yang--Mills and gravitational couplings. See also \cite{Buchbinder:2006eq} for a recent work 
concerning higher-spin vertices, including a discussion about the $AdS_n$ background. 

The Poincar\'e-invariant, local, non-Abelian consistent cross-interactions between spin-2 and 
spin-3 gauge fields in flat space remained to be analyzed in an exhaustive way and 
without any prejudice on the form of the interactions. In particular, we impose no 
upper limit on the number of derivatives appearing in the non-Abelian consistent vertex, 
apart that it should be finite in order that locality be preserved. 

The advantage of the cohomological 
method \cite{Barnich:1993vg,Henneaux:1997bm} which we use is that it enables one to classify and explicitly write down the consistent, nontrivial cubic vertices, 
without any other assumptions than locality 
and perturbative nature of the deformations. This method is also compatible with manifest 
Poincar\'e and gauge invariances, which is of great importance in the search for a 
possible geometrical interpretation of the higher-spin interactions. 
Since we have access to all the possible local, perturbative deformations of the 
gauge algebra and gauge transformations giving rise to nontrivial consistent cubic vertices, 
it can be hoped that the deformed gauge transformations provide crucial information on
a possible underlying nonlinear higher-spin geometry in flat space. Such a 
geometrical picture would in turn guide us toward a full nonlinear consistent Lagrangian. 

Similarly to the self-interacting totally symmetric spin-2 and spin-3 cases \cite{Boulanger:2000rq,Boulanger:2000ni,Bekaert:2005jf,Boulanger:2005br}, 
we first classify the possible first-order deformations of the gauge algebra and 
then determine which of these deformations give rise to nontrivial, consistent  
vertices. 
It turns out that only two parity-invariant algebra-deforming candidates satisfy this 
strong requirement.
Interestingly enough, we find that, in order for the first candidate to induce a  
Poincar\'e-invariant nontrivial vertex, the (colored) spin-2 massless fields must 
react to the spin-3 field through a diffeomorphism-like transformation along the
spin-3 gauge parameter, similarly to the way a spin-1 field reacts to a gravitational 
background via its Lie derivative along the diffeomorphism vector. 
Associated with the second algebra-deforming candidate is a gauge transformation of the spin-3 
field along its own gauge parameter, but involving the linearized Riemann tensor for
the spin-2 field. The first algebra-deforming candidate corresponds to the $3-2-2$
covariant vertex mentioned in \cite{Berends:1984rq,Berends:1985xx}, 
whereas the second algebra-deforming candidate gives rise to a nontrivial consistent 
$2-3-3$ vertex which had previously not been written before, to our knowledge. 

Our results therefore strengthen and complete those previously found in 
\cite{AL,Berends:1984rq,Berends:1985xx}. 
In particular, we recover in a simple way that both minimal and non-minimal couplings 
of spin-3 gauge fields to dynamical gravity in flat space are inconsistent \cite{AL}. 

In the work \cite{Berends:1984rq}, consistent and covariant cubic couplings
of the kind $s_1-s_2-s_2$ were obtained, for the values of 
$s_1$ and $s_2$ indicated in Table \ref{T1}. 
\begin{table}[!ht]
\centering
\begin{tabular}{c |c c c c c c c }
 ${\downarrow}_{s_1} \quad {\rightarrow}^{s_2}$ 
  & $0$  & $\frac{1}{2}$  & $1$  & $\frac{3}{2}$ &  $2$
& $\frac{5}{2}$ & $3$  \\ \hline\hline
$ 0\qquad $   & $\times$  & $\times$  & $\times$ & $\times$ & $\times$ &  &  \\
\hline
$ 1\qquad $   &$\times$  & $\times$  &$\times$ & $\times$ & $\times$ &  &  \\
\hline
$ 2\qquad $   &$\times$  & $\times$  &$\times$ & $\times$ & $\times$ & $\times$ &  \\
\hline
$ 3\qquad $   &$\times$  & $\times$  & $\times$ & $\times$ & $\times$ & $\times$ & $\times$ \\
\hline
$n\qquad$ & $\times$
\\ \hline
\end{tabular}
\caption{\it  $s_1-s_2-s_2$ covariant vertices obtained in \cite{Berends:1984rq}.
\label{T1}}
\end{table} 
Of course, some of the vertices were already known before, 
like for example in the cases $1-1-1$, $2-2-2$ and $2-\frac{3}{2}-\frac{3}{2}$ 
corresponding to Yang--Mills, Einstein--Hilbert and ordinary supergravity theories.
There is a class of cross-interactions $s_1-s_2-s_2$ for which the cubic vertices could 
easily been written. 
This class corresponds to the ``Bell--Robinson'' line $s_1=2s_2$ 
and below this line $s_1>2s_2$ \cite{Berends:1985xx} 
(see \cite{Deser:1990bk} in the particular $s_1=4=2s_2$ case). 
In the aforementioned region $s_1\geqslant 2 s_2$, 
the gauge algebra remains Abelian although the
gauge transformations for the spin-$s_2$ field are deformed at first order in a 
coupling constant. The reason is that the first-order deformation of the free 
spin-$s_2$ gauge transformations involve the spin-$s_2$ field only through its 
gauge-invariant Weinberg--de Wit--Freedman field-strength 
\cite{deWit:1979pe,Weinberg:1965rz}\footnote{Note that one can write down
higher-derivative Born--Infeld-like consistent cubic interactions involving 
only gauge-invariant field-strength tensors \cite{Damour:1987fp}. 
However, these interactions deform neither the gauge algebra nor the gauge
transformations. They are not considered in the present work but are 
accounted for in the powerful light-cone approach presented in \cite{Metsaev:2005ar}.}. 
Although they do not lead to non-Abelian gauge algebras, it is interesting that 
the cubic interactions on and below the Bell--Robinson line
(\textit{i.e.} for $s_1\geqslant 2s_2$) have the form 
``spin-$s_1$ field times current $J$'' where $J$ is quadratic in 
the spin-$s_2$ field-strength \cite{Deser:1990bk,Berends:1985xx} and is conserved 
on the spin-$s_2$ shell.  
Even more interestingly, these currents can be obtained from some 
global invariances of the free theory by a Noether-like procedure, provided the constant
parameters associated with these rigid symmetries be replaced by the gauge parameters
of the spin-$s_1$ field (also internal indices must be treated appropriately)  
\cite{Deser:1990bk,Berends:1985xx}. 

In the present paper, we re-derive the non-Abelian $3-2-2$ cubic vertex mentioned in 
\cite{Berends:1984rq}, show that it is inconsistent when pushed up to second order in 
the deformation parameter and obtain a consistent $2-3-3$ vertex which passes the
second-order consistency test where the former Berends--Burgers-van Dam vertex fails. 
Moreover, at the level of the Jacobi identity at second order in the coupling constant, 
we show that the latter $2-3-3$ covariant vertex is compatible with the spin-3 
self-coupling written in \cite{Berends:1984wp}. 
Also, even though the $3-2-2$ vertex stands above the Bell--Robinson line $s_1=2s_2$, 
we show that it can be seen as partially resulting from the gauging of the global 
symmetries discussed in \cite{Berends:1985xx}. Of course, this vertex truly
deforms the gauge algebra and consequently the coupling cannot be written in 
the simple form outlined before. The gauge transformations for both spin-2 and
spin-3 fields are nontrivially deformed.

The plan of the paper is as follows. 
In the next section we recall some basic facts on the free spin-2 and spin-3 
gauge theories and on the BRST-antifield formalism used throughout the text. 
Section \ref{sec:cou} gathers together some BRST-cohomological results 
that are needed. Section \ref{sec:dfdf} contains most 
of our computations and results concerning the first-order consistent couplings 
between spin-2 and spin-3 massless fields. In Section \ref{sec:further} we present
the constraints that are imposed on the first-order deformations by second-order 
consistency conditions.  
We also discuss the links between the first-order gauge transformations presented 
in Section \ref{sec:dfdf} and some results of \cite{Deser:1990bk,Berends:1985xx}. 
Finally, our conclusions and perspectives are given in
Section \ref{sec:concl}. 
The first appendix contains a technical BRST-cohomological result. 
The complete expressions for the first-order vertices are displayed
in the second appendix.

%*********************************%%%%%%%%%%%%%%%
\section{Free theory and BRST settings}
\label{sec:Settings}
%*********************************%%%%%%%%%%%%%%%

\subsection{Free Theory}
%---------------------------
The action for a collection $\{h_{\mu \nu}^a\}$ 
of $M$ {\it non-interacting}, massless spin-2 fields in spacetime dimension
$n$ ($\mu, \nu = 0, \cdots, n - 1$) 
is (equivalent to) the sum of $M$ separate Pauli-Fierz actions, 
namely
\beq 
\label{startingpoint} 
S^2_0[h_{\mu \nu}^a] &=& \sum_{a = 1}^M \int \delta_{ab}\,\left[ 
-\frac{1}{2}\left(\pa_{\m}{h^a}_{\n\r}\right)\left(\pa^{\m}{h}^{b\n\r}\right) 
+\left(\pa_{\m}{h^a}^{\m}_{~\n}\right)\left(\pa_{\r}{h}^{b\r\n}\right)\right.
\nonumber \\ 
&&\left. -\left(\pa_{\n}{h^a}^{\m}_{~\m}\right)\left(\pa_{\r}{h}^{b\r\n}\right) 
+\frac{1}{2}\left(\pa_{\m}{h^a}^{\n}_{~\n}\right) 
\left(\pa^{\m}{h}^{b\r}_{~\r}\right)\right] d^n x \, , \; \; n>2. 
\label{action}
\eeq 
The lower-case Latin indices are internal indices taking $M$ values. 
They are raised and lowered with the Kronecker delta's $\d^{ab}$ and $\d_{ab}$.
The Greek indices are space-time indices taking $n$ values, which are
lowered (resp. raised) with the ``mostly plus" Minkowski metric $\eta_{\m\n}$ (resp. $\eta^{\m\n}$).
The action (\ref{startingpoint})
is invariant under the following linear gauge transformations, 
\beq 
\delta_\epsilon h^a_{\mu \nu} = \partial_\mu \epsilon_\nu^a 
+ \partial_\nu \epsilon_\mu^a 
\label{freegauge} 
\eeq 
where the $\epsilon_\nu^a$ are $n \times M$ arbitrary, independent functions.
These transformations are Abelian and irreducible.  
The equations of motion are 
\begin{eqnarray} 
 \frac{\d S^2_0}{\d h^a_{\mu \nu}} =  - 2 H_a^{\mu \nu} = 0 
\nonumber
\end{eqnarray} 
where $H^a_{\mu \nu}$ is 
the linearized Einstein tensor,
\beq 
H^a_{\mu \nu} =  K^a_{\mu \nu} - \frac{1}{2} K^a \eta_{\mu \nu}\,. 
\nonumber
\eeq 
Here, $K^a_{\a \b \m \n}$  is the linearized Riemann tensor, 
\begin{eqnarray} 
K^{a}_{\a\b\m\n}=- \frac{1}{2}
(\pa_{\a\m}h_{\b\n}^{a}+\pa_{\b\n}h_{\a\m}^{a} 
-\pa_{\a\n}h_{\b\m}^{a} 
-\pa_{\b\m}h_{\a\n}^{a})\,, 
\nonumber
\end{eqnarray}
$K^a_{\mu \nu}$ is the linearized Ricci tensor, 
\begin{eqnarray} 
K^a_{\mu \nu} = K^{a \a}_{~~ \,\m \a \n} = - \frac{1}{2}
(\Box h^a_{\mu \nu} + \cdots)\,,
\nonumber
\end{eqnarray}
and $K^a$ is the linearized scalar curvature, 
$K^a = \eta^{\mu \nu}  K^a_{\mu \nu}\,$.  
The Noether identities expressing the invariance of the free 
action (\ref{action}) under (\ref{freegauge}) are 
\beq 
\partial_\nu H^{a \mu \nu} = 0 
\label{linBianchi} 
\eeq 
(linearized Bianchi identities).  
The gauge symmetry removes unwanted unphysical states. 
\vspace*{.5cm}

The local action for a collection
$\{h^A_{\m\n\r}\}$ of $N$ non-interacting totally symmetric
massless spin-3 gauge fields in flat spacetime is
\cite{Fronsdal:1978rb}
\begin{eqnarray}
S^3_0[h^A_{\m\n\r}] = \sum_{A=1}^N \int 
                \delta_{AB}&\Big[& -\frac{1}{2}\,\pa_{\s}h^A_{\m\n\r}\pa^{\s}h^{B\m\n\r} +
                        \frac{3}{2}\,\pa^{\m}h^A_{\m\r\s}\pa_{\n}h^{B\n\r\s} +
          \nonumber \\
               && \frac{3}{2}\,\pa_{\m}h^A_{\n}\pa^{\m}h^{B\n} +
                  \frac{3}{4}\,\pa_{\m}h^{A\m}\pa_{\n}h^{B\n} -
                        3 \,\pa_{\m}h^A_{\n}\pa_{\r}h^{B\r\m\n} \,\,\,\Big] d^n x\,,
\label{freeaction}
\end{eqnarray}
where $h^A_{\m}=\eta^{\n\r}h^A_{\m\n\r}\,$.
The upper-case Latin indices are internal indices taking $N$ values. They are raised
and lowered with the Kronecker delta's $\d^{AB}$ and $\d_{AB}$.
The action (\ref{freeaction}) is invariant under the gauge
transformations
\begin{eqnarray}
        \delta_{\l}h^A_{\m\n\r} = 3 \,\pa^{}_{(\m}\l^A_{\n\r)}\,,\quad 
        \eta^{\m\n}\l^A_{\m\n}=  0\,,
\label{freegauge3}
\end{eqnarray}
where the gauge parameters $\l^A_{\n\r}$ are symmetric and traceless. 
Curved (resp. square) brackets on spacetime indices denote strength-one complete symmetrization
(resp. antisymmetrization) of the indices.
The gauge transformations (\ref{freegauge3}) are Abelian and irreducible.
The field equations read
\begin{eqnarray}
        \frac{\d S^3_0}{\d h^A_{\m\n\r}} =  G^{\m\n\r}_A = 0\,,
        \label{eom}
\end{eqnarray}
where
\begin{eqnarray}
G^{A}_{\m\n\r}=F^{A}_{\m\n\r} -\frac{3}{2}\,
\eta^{}_{(\m\n}F^A_{\r)} 
\label{einstein}
\end{eqnarray}
is the ``Einstein" tensor and $F^{A}_{\m\n\r}$ the Fronsdal (or
``Ricci") tensor
\begin{eqnarray}
        F^{A}_{\m\n\r} =
        \Box h^{A}_{\m\n\r} - 3 \,\pa^{\s}\pa^{}_{(\m}h^A_{\n\r)\s} + 3 \, \pa^{}_{(\m}\pa^{}_{\n}h_{\r)}^A\,.
\label{Fronsdal}
\end{eqnarray}
The Fronsdal tensor is gauge invariant thanks to the tracelessness
of the gauge parameters. 
Because we have $\d_{\l}S_0[h^A_{\m\n\r}]= 0\,$ for the gauge transformations 
(\ref{freegauge3}), the Einstein tensor $G^{A}_{\m\n\r}$ satisfies the Noether 
identities
\begin{eqnarray}
        \pa^{\r}G^{A}_{\m\n\r}-\frac{1}{n}\,\eta_{\m\n}\pa^{\r}G^{A}_{\r}=  0
        \quad \quad (G^{A}_{\r} = \eta^{\m\n} G^{A}_{\m\n\r})
\label{Noether}
\end{eqnarray}
related to the symmetries of the gauge parameters $\l^A_{\m\n}\,$; in other words,
the l.h.s. of (\ref{Noether}) is symmetric and traceless.

An important object is the Weinberg--de Wit--Freedman (or ``Riemann") spin-3 tensor
\cite{Weinberg:1965rz,deWit:1979pe,Damour:1987vm}
\begin{eqnarray}
        K^A_{\a\m|\b\n|\g\r}= 8 \pa^{}_{[\g} \pa^{}_{[\b}\pa^{}_{[\a}h^A_{\m]\n]\r]}
\nonumber
\end{eqnarray}
which is antisymmetric in $\a\m\,$, $\b\n\,$, $\g\r$ and invariant under gauge transformations
(\ref{freegauge3}), where the gauge parameters $\lambda^A_{\m\n}$ are however {\textit{not}} necessarily traceless.
Its importance, apart from gauge invariance with unconstrained gauge parameters, stems from the fact that the field equations ({\ref{eom}}) are equivalent
to the following equations
\begin{eqnarray}
        \eta^{\a\b} K^A_{\a\m|\b\n|\g\r} = 0\,,
\nonumber
\end{eqnarray}
after a partial gauge fixing. 
This was proved in the work \cite{BB} by combining various former
results \cite{Damour:1987vm,DVH,FS}. 
See \cite{Bekaert:2006ix} for more details and for the arbitrary mixed-symmetry case.

%------------------------------------------
\subsection{BRST spectrum and differential}
\label{BRSTspectrum}
%------------------------------------------
%
According to the general rules of the BRST-antifield formalism, the field spectrum
consists of the fields $\{h^a_{\m\n},h^A_{\m\n\r}\}\,$, the ghosts $\{C^a_\m,C^A_{\m\n}\}$, the antifields $\{h^{*\m\n}_a,h^{*\m\n\r}_A\}$ and the ghost antifields $\{C^{*\m}_a,C^{*\m\n}_A\}$. The set of fields and ghosts will sometimes be collectively denoted by $\Phi^I$, whereas 
the associated set of antifields will be denoted by $\Phi^*_I\,$. 

The BRST differential $s$ of  the free theory  
$S_0[h^a,h^A]=S^2_0[h^a_{\m\n}]+S^3_0[h^A_{\m\n\r}]$ 
is generated by the functional
\begin{eqnarray}
W_0 =  S_0[h^a,h^A]\;+ \int ( 
\, 2\,h^{* \a\b}_{a} \partial_\a C^a_{\b} + 
3\, h^{*\m\n\r}_A \, \pa_{\m} C_{\n\r}^A )\; d^n x\,. \nonumber
\end{eqnarray}
More precisely, $W_0$ is the generator of the BRST differential $s$ of the free theory through
\begin{eqnarray}
        s A = (W_0, A)\,, \nonumber
\end{eqnarray}
where the antibracket $(~,~)$ is defined by
\begin{eqnarray}
(A,B)=\frac{\d^R A}{\d \Phi^I}\frac{\d^L B}{\d \Phi^*_I} -
 \frac{\d^R A}{\d \Phi^*_I}\frac{\d^L B}{\d \Phi^I}\,,
 \nonumber
\end{eqnarray}
using the condensed de Witt notation in which a summation over a repeated index 
also implies an integration over spacetime variables.  

The functional $W_0$ is a solution of the \emph{master equation}
\begin{eqnarray}
        (W_0,W_0)=0\,.
        \nonumber
\end{eqnarray}
In the theory at hand, the BRST-differential $s$ decomposes into $s=\g + \d \,$.
The first piece $\g\,$, the differential along the gauge orbits, is associated with the $\mathbb{N}$-grading called pureghost number ($puregh$) and increases it by one unit, whereas the Koszul-Tate differential $\d$ decreases the grading called antighost number ($antigh$) by one unit.
The differential $s$ increases the $\mathbb{Z}$-grading called ghost number ($gh$) by one unit.
Furthermore, the ghost, antighost and pureghost gradings are not
independent. We have the relation
\begin{eqnarray}
        gh = puregh - antigh\,.
        \nonumber
\end{eqnarray}
The pureghost number, antighost number, ghost number and grassmannian parity of the various fields are displayed in Table \ref{table1}.
\begin{table}[!ht]
\centering
\begin{tabular}{|c|c|c|c|c|}
\hline Z  & $puregh(Z)$  & $antigh(Z)$  & $gh(Z)$  & parity (mod $2$)\\ \hline
$\{h^a_{\m\n},h^A_{\m\n\r}\}$   &$0$  & $0$  &$0$ &$0$ \\
$\{C^a_{\m},C^A_{\m\n}\}$ & $1$ & $0$ & $1$ & $1$ \\
$\{h^{*\m\n}_a,h^{*\m\n\r}_A\}$ & $0$& $1$ & $-1$ & $1$ \\
$\{C^{*\m}_a,C^{*\m\n}_A\}$ & $0$ & $2$ & $-2$ & $0$ \\
\hline
\end{tabular}
\caption{\it pureghost number, antighost number, ghost number
and parity of the (anti)fields.\label{table1}}
\end{table}
The action of the differentials  $\delta$ and $\gamma$ gives zero on all the
fields of the formalism except in the few following cases:
\begin{eqnarray}
&\d h^{*\m\n\r}_A =G^{\m\n\r}_A\,,\qquad \d h^{*\m\n}_a = -2H^{\m\n}_a\,, &
\nonumber \\
&\d C^{*\m\n}_A = -3 ( \pa_{\r}h^{*\m\n\r}_A - \frac{1}{n}\eta^{\m\n}\pa_{\r}h^{*\r}_A )\,,
\qquad \d C_a^{*\m}=-2\pa_{\n}h_a^{*\n\m}\,, &
\nonumber \\
&\gamma h^A_{\m\n\r} = 3\,\pa^{}_{(\m}C_{\n\r)}^A\,,\qquad
\gamma h^a_{\m\n} = 2\,\pa^{}_{(\m}C_{\n)}^a\,. &
\nonumber 
\end{eqnarray}
More details about the antifield formalism for spin-2 and spin-3 can be found in \cite{Boulanger:2000rq,Bekaert:2005jf}.
%
%------------------------------------------
\subsection{BRST deformations}
\label{deformation}
%---------------------------------
%
As shown in \cite{Barnich:1993vg}, the Noether procedure can be
reformulated within a BRST-cohomological framework. Any
consistent deformation of the gauge theory corresponds to a
solution $$W=W_0+g W_1+g^2W_2+\co(g^3)$$ of the deformed master
equation $(W,W)=0$. Consequently, the first-order nontrivial
consistent local deformations $W_1=\int a^{n,\,0}$ are in
one-to-one correspondence with elements of the cohomology
$H^{n,\,0}(s \vert\, d)$ of the zeroth order BRST differential
$s=(W_0,\cdot)$ modulo the total derivative $d\,$, in maximum
form-degree $n$ and in ghost number $0\,$. That is, one must
compute the general solution of the cocycle condition
\begin{eqnarray}
        s a^{n,\,0} + db^{n-1,1} =0\,,
        \label{coc}
\end{eqnarray}
where $a^{n,\,0}$ is a top-form of ghost number zero and
$b^{n-1,1}$ a $(n-1)$-form of ghost number one, with the
understanding that two solutions of (\ref{coc}) that differ by a
trivial solution should be identified
\begin{eqnarray}
        a^{n,\,0}\sim a^{n,\,0} + s p^{n,-1}  + dq^{n-1,\,0} \,,\nonumber
\end{eqnarray}
as they define the same interactions up to field redefinitions.
The cocycles and coboundaries $a,b,p,q,\ldots\,$ are local forms of
the field variables, including ghosts and antifields (those are forms acting on the jet space $J_k$, the vectorial space generated by the fields and a finite number $k$ of their derivatives).

The corresponding second-order interactions $W_2$ must satisfy the consistency condition
\begin{eqnarray}
s  W_2=-\frac{1}{2} (W_1,W_1)\,.
\label{secor}
\end{eqnarray}
This condition is controlled by the local BRST cohomology group $H^{n,1}(s\vert d)$.

%*********************************%%%%%%%%%%%%%%%
\section{Cohomological results}
\label{sec:cou}
%*********************************%%%%%%%%%%%%%%%

%-----------------------------------------------
\subsection{Cohomology of $\g$}
\label{hogamma}
%-----------------------------------------------

In the context of local free theories in Minkowski space for
massless spin-$s$ gauge fields represented by totally symmetric
(and double traceless when $s>3$) rank $s$ tensors, the groups
$H^*(\g)$ have been calculated in \cite{Bekaert:2005ka}.
When a sum of several such theories for different spins is considered, the cohomology is the direct product of the cohomologies of the different theories. We will prove it only in the case of a sum of spin-2 and spin-3 Fronsdal theories, but the proof can straightforwardly be 
extended.

\begin{prop}\label{Hgamma} The cohomology of $\g$ is
isomorphic to the space of functions depending on
\begin{itemize}
  \item the antifields $\{ h^{*\m\n}_a,~h^{*\m\n\r}_A,~C^{*\m}_a,~C^{*\m\n}_A \}$ 
  and their derivatives, denoted by ${[}\Phi^{*}_I{]}\,$,
  \item the curvatures and their derivatives 
  $[K^a_{\a\m\b\n}]\,$, $[K^A_{\a\m|\b\n|\g\r}]\,$,
  \item the symmetrized derivatives $\pa^{}_{(\s_1}\ldots\pa^{}_{\s_k}F^A_{\m\n\r)}$ of the 
  Fronsdal tensor,
  \item the ghosts $C^a_{\mu}$ and their antisymmetrized first-order derivatives 
  $\pa_{[\m}C^a_{\n]}$, 
  the ghosts $C_{\m\n}^A$ and the traceless parts of $\pa^{}_{[\m}C_{\n]\r}^A$ 
  and\footnote{A coma denotes a partial derivative, e.g.
  $\Phi^I_{,\m}=  \partial_{\m}\Phi^I\,$. } $\pa^{}_{[\m}C_{\n][\r,\s]}^A$.
\end{itemize} 
Thus, identifying with zero any $\gamma$-exact term in $H(\gamma)$, we have   
$$ \g f=0 $$
if and only if $$f=
f\left([\Phi^{*}_I],[K^A_{\a\m|\b\n|\g\r}],[K^a_{\a\m\b\n}],\{F^A_{\m\n\r}\},
        C^a_{\n},\pa_{[\m}C^a_{\n]},C_{\m\n}^A, \widehat{T}^A_{\a\m\vert\n}, \widehat{U}^A_{\a\m\vert\b\n}
     \right)$$
where $\{F^A_{\m\n\r}\}$ stands for the completely symmetrized
derivatives $\pa^{}_{(\s_1}\ldots\pa^{}_{\s_k}F^A_{\m\n\r)}$ of
the Fronsdal tensor, while $\widehat{T}^A_{\r\m\vert\n}$ denotes
the traceless part of $T^A_{\r\m\vert\n}=\pa^{}_{[\r}C_{\m]\n}^A$ and $\widehat{U}^A_{\r\m\vert\s\n}$ the
traceless part of ${U}^A_{\r\m\vert\s\n}=\pa^{}_{[\r}C_{\m][\n,\s]}^A\,$.
\end{prop}

%\vspace{.3cm}

Let $\{\o^I\}$ be a basis of the space of polynomials in the
$C_\m^a$, $\pa_{[\mu}C_{\nu]}^a$, $C_{\m\n}^A$, $\widehat{T}^A_{\a\m\vert\n}$ and $\widehat{U}^A_{\a\m\vert\b\n}$
(since these variables anticommute, this space is finite-dimensional).
If a local form $a$ is $\gamma$-closed, we have
\begin{eqnarray}\label{alom}
        \g a = 0 \;\Rightarrow\; a \,=\,
        \a_J([\Phi^{*}_I],[K^a],[K^A],\{F^A\})\,
        \o^J(C_\m^a, \pa_{[\mu}C_{\nu]}^a, 
        C_{\m\n}^A,\widehat{T}^A_{\a\m\vert\n},\widehat{U}^A_{\a\m\vert\b\n}) + \g b\,.
\end{eqnarray}
If $a$ has a fixed, finite ghost number, then $a$ can only contain
a finite number of antifields. Moreover, since the
{\textit{local}} form $a$ possesses a finite number of
derivatives, we find that the $\a_J$ are polynomials. Such a
polynomial $\a_J([\Phi^{*}_I],[K^a],[K^A],\{F^A\})$ will be called  an
{\textit{invariant polynomial}}.
The proof of Proposition \ref{Hgamma} is given in appendix \ref{SergeA}.
\vspace{.3cm}

\noindent {\textbf{Remark}:} Because of the Damour-Deser identity
\cite{Damour:1987vm}
$$\eta^{\a\b}K^A_{\a\m|\b\n|\g\r}=  2\, \pa_{[\g}F^A_{\r]\m\n}\,,$$ 
the derivatives of the Fronsdal tensor $F^A$ are not all independent  
of the curvature tensor $K^A\,$. 
This is why, in Proposition \ref{Hgamma}, the completely
symmetrized derivatives of $F^A$ appear, together with all the
derivatives of the curvature $K^A\,$. However, from now on, we will
assume that every time the trace $\eta^{\a\b}K^A_{\a\m|\b\n|\g\r}$
appears, we substitute $2\pa_{[\g}F^A_{\r]\m\n}$ for it. 
With this convention, we can write $\a_J([\Phi^{*}_I],[K^a],[K^A],[F^A])$ 
instead of the inconvenient notation $\a_J([\Phi^{i*}],[K^a],[K^A],\{F\})\,$.

%-------------------------------------------------------
\subsection{Invariant Poincar\'e lemma}
\label{invPlemma}
%--------------------------------------------------------

We shall need several standard results on the cohomology of $d$ in
the space of invariant polynomials.
\begin{prop}\label{2.2}
In form degree less than $n$ and in antifield number strictly greater than $0$,
the cohomology of $d$ is trivial in the space of invariant
polynomials.
That is to say, if $\a$ is an invariant polynomial, the equation
$d \a = 0$ with $antigh(\a) > 0$ implies
$ \a = d \b$ where $\b$ is also an invariant polynomial.
\end{prop}
\noindent The latter property is rather generic for gauge theories
(see e.g. Ref. \cite{Boulanger:2000rq} for a proof), as well as
the following:

\begin{prop}\label{csq}
If $a$ has strictly positive antifield number, then the equation
$\gamma a + d b = 0$ is equivalent, up to trivial redefinitions,
to $\gamma a = 0$. More precisely, one can always add $d$-exact
terms to $a$ and get a cocycle $a' = a  + d c$ of $\gamma$, such
that $\g a'= 0$.
\end{prop}
\vspace*{.2cm}

\noindent\textbf{Proof: } See e.g. \cite{Bekaert:2005jf}. 

%---------------------------------------------------------------------
\subsection{Cohomology of $\d$ modulo $d\,$: $H^n_k(\d \vert\, d)$}
\label{Characteristiccohomology}
%----------------------------------------------------------------------

In this section, we review the local Koszul-Tate cohomology
groups in top form-degree and antighost numbers $k\geqslant 2\,$.
The group $H^n_1(\d \vert\, d)$ describes the infinitely many
conserved currents and will not be studied here. 
\vspace*{.2cm}

Let us first recall a general result (Theorem 9.1 in \cite{Barnich:1994db}).

\begin{prop}\label{usefll}
For a linear gauge theory of reducibility order $r$,
\begin{eqnarray}
H_p^n(\d \vert\, d)=0\; for\; p>r+2\,. \nonumber
\end{eqnarray}
\end{prop}
Since the theory at hand has no reducibility, we are left with the computation of
$H_2^n(\d \vert\, d)\,$. 
The cohomology $H_2^n(\d \vert\, d)$ is given by the following theorem.

\begin{prop}\label{H2}
A complete set of representatives of $H^n_2(\d\vert d)$ is given by the antifields
$\{C_a^{*\m},C_A^{*\m\n}\}$, up to explicitly $x$-dependent terms. In detail,
\begin{eqnarray}
        \left\{
        \begin{array}{ll}
         \delta a^n_2 + d b^{n-1}_1 = 0
        \nonumber \\
         \quad a^n_2 \sim  a^n_2 + \delta c^n_3 + d c^{n-1}_2
         \end{array}\right.
           \Longleftrightarrow 
                a^n_2 = \left[\l^a_\mu(x)C^{*\mu}_a
        +L^A_{\m\n}(x)C_A^{*\m\n}\right]d^n x + \delta b^n_3 + d b_2^{n-1}
\end{eqnarray}
where $\l^a_\mu(x)=a_\mu^a+\ldots$ is a degree-1 polynomial in $x^\nu$ and 
$L^A_{\m\n}(x)=\l^A_{\mu\nu}+\ldots$ is a degree-2 polynomial.
\end{prop} 
\noindent The coefficients of these polynomials have definite symmetry properties that we will not recall here. The complete analysis can be found in 
\cite{Boulanger:2000rq,Bekaert:2005jf}. See also \cite{Barnich:2005bn,Nazim}. 
{}From the requirement of Poincar\'e invariance, explicit dependence in the coordinates is 
forbidden and we will only consider the constant terms $a_\mu^a$ and $\l^A_{\mu\nu}$
in the expansions of $\l^a_\mu(x)$ and $L^A_{\m\n}(x)$.

\noindent The most general $n$-form in $antigh\ 2$ is $a=(f_\mu^a 
C^{*\mu}_a+\widehat{f}_{\mu\nu}^A C^{*\mu\nu}_A)d^n x + \Phi + \d b + dc$, where $\Phi$ is 
quadratic in the $antigh$-$1$ antifields. 
If one applies $\d$, the $\delta$-exact term vanishes 
and $\d\Phi + df \approx 0$. So, if $a\in H_2^n(\d|d)$, the weak equality 
$(-2f_\mu^a\6_\nu h^{*\mu\nu}-3\widehat{f}_{\mu\nu}^A\6_\r h^{*\mu\nu\r}_A)d^n x\approx dv$ 
is obtained. 
Finally, by 
applying variational derivatives with respect to $h^{*\mu\nu}_a$ and $h^{*\m\n\r}_A$, the
two weak equalities 
$\6_{(\mu}f^a_{\nu)}\approx 0$ and $\6_{(\mu}\widehat{f}^A_{\n\r)}\approx 0$ are obtained. These are both on-shell Killing equations for the individual spin-2 and spin-3 cases. 
Each equation of the type $\6_{(\mu_1}\widehat{f}_{\mu_2...\mu_s)}\approx 0$ provides 
$H_2^n(\d|d)$ for the pure spin-$s$ case, the solutions of which have been given in 
Ref.~\cite{Bekaert:2005ka} (see also \cite{Barnich:2005bn}). 
This is because those solutions are $\d$-closed modulo $d$ and because $\Phi$ obeying 
$\d\Phi+d c=0$ is a trivial cocycle \cite{Barnich:1994mt}. 
The spin-$2$ case under consideration was already written in 
 Ref.~\cite{Boulanger:2000rq} and the spin-$3$ case was written in \cite{Bekaert:2005jf,Nazim}. 
In any mixed case, the different equations for the different spins will have to be satisfied and $H^n_2(\d|d)$ is then the direct sum of the individual cases.
\vspace*{.3cm}

We have studied above the cohomology of $\delta$ modulo $d$ in the
space of arbitary local functions of the fields, the antifields, and their derivatives.  
One can also study $H^n_k(\delta \vert d)$ in the space of invariant polynomials in these variables. The above theorems remain unchanged in this
space, {\emph{i.e.}} we have the
\begin{prop}\label{invKo}
The invariant cohomology $H^{n}_k(\delta\vert d,H_0(\g))$ is trivial 
in antighost number $k>2\,$. In antighost number $k=2$, we have
the isomorphism $H^{n}_2(\delta\vert d,H_0(\g))\cong H^{n}_2(\delta\vert d)\,$.  
\end{prop}
This very nontrivial property is crucial for the computation of 
$H^{n,0}(s \vert\, d)$. It has been proved for the spin-2 case in \cite{Boulanger:2000rq} and for the spin-3 case in \cite{Bekaert:2005jf} for $n>3$, and \cite{Boulanger:2005br} for $n=3\,$. In the mixed spin-2--spin-3 case, the proof goes along the same lines. It has to be checked that, in a coboundary in form degree $n$ : $a_k=\d b_{k+1}+ \6_\mu j^{\mu}_k$ (in dual notation), if $a_k$ is invariant, then $b_{k+1}$ and $j^{\mu}_k$ can be chosen as being invariant. This is done by reconstructing $a_k$ from its variational derivatives with respect to the different fields and antifields (see \cite{Boulanger:2000rq}, Lemma A.2 and \cite{Bekaert:2005jf}, section 4.6.2). The considerations made for the spin-2 and spin-3 derivatives hold independently here, and $a_k$ can be reconstructed in an invariant way with no further problems.

%---------------------------
\subsection{Definition of the $D$-degree}
\label{Ddeg}
%--------------------------

\noindent \textbf{Definition (differential $D$)}: 
The action of the differential $D$ on the fields, the antifields and all their derivatives
is the same as the action of the total
derivative $d$, but its action on the ghosts is given by:
\begin{eqnarray}
D C^A_{\m\n} &=& {\frac {4}{3}} \, d x^{\a}\, {\widehat{T}}^A_{\a(\mu\vert\nu )}\,,
\nonumber \\
D  {\widehat{T}}^A_{\m\a\vert\b} &=& d x^{\r} \, {\widehat{U}}^A_{\m\a\vert\r\b}\,,
\nonumber \\
D(\partial _{\rho _1 \ldots \rho _t} C^A_{\mu\nu}) &=& 0 ~ {\rm \ if ~}~ t\geqslant 2\,, \nonumber \\ 
D C_{\mu}^a &=& dx^{\nu} \partial_{[\nu} C^a_{\mu]} \,, \nonumber \\
D (\partial _{\rho _1 \ldots \rho _v} C^a_{\mu}) &=& 0 ~ {\rm \ if ~}~ v\geqslant 1\,.
\nonumber
\end{eqnarray}
The above definitions follow from
\begin{eqnarray}
        \partial_{\m} C^a_{\n} &=& \frac{1}{2}\,
        ( \gamma h^a_{\m\n} ) + \partial_{[\mu} C^a_{\n]}  \,,
        \nonumber \\
        \pa_{\a}C^A_{\m\n} &=& \frac{1}{3}
        (\gamma h^A_{\a\m\n})+\frac{4}{3}}\,T^A_{\a(\m\vert\n)\,,
        \nonumber \\
        \pa_{\r}T^A_{\m\a\vert\b} &=& 
        -\frac{1}{2}\,\g(\pa_{[\a}h^A_{\m]\b\r})+U^A_{\m\a\vert\r\b}\,,
        \nonumber \\
        \pa_{\r}U^A_{\m\a\vert\n\b} &=& \frac{1}{3}\, \g (\pa_{[\m}h^A_{\a]\r[\b,\n]})\,.
\nonumber
\end{eqnarray}
The operator $D$ thus coincides with $d$ up to $\gamma$-exact terms.

It follows from the definitions that $D\omega^J = A^J{}_I
\omega^I$ for some constant matrix $A^J{}_I$ that involves $dx^\m$
only. It is also convenient to introduce a new grading.
\vspace{2mm}

\noindent\textbf{Definition ($D$-degree)}: 
The number of ${\widehat{T}}^A_{\a\m|\n}$'s and 
$\partial_{[\mu}C^a_{\nu]}$'s plus twice the number of 
${\widehat{U}}^A_{\a\m|\b\n}$'s is called the $D$-degree. It is
bounded because there is a finite number 
of $ \partial_{[\mu}C^a_{\nu]}$'s, ${\widehat{T}}^A_{\a\m|\n}$'s and ${\widehat{U}}^A_{\a\m|\b\n}$'s which are anticommuting.
The operator $D$ splits as the sum of an operator $D_1$ that raises 
the $D$-degree by one unit and an operator $D_0$ that leaves it unchanged. 
$D_0$ has the same action as $d$ on the fields, the antifields and
all their derivatives, and gives $0$ when acting on the ghosts. 
$D_1$ gives $0$ when acting on all the variables but the ghosts on 
which it reproduces the action of $D$.

%****************************************
\section{First-order consistent deformations}
\label{sec:dfdf}
%****************************************

As recalled in Section \ref{deformation}, nontrivial
consistent interactions are in one-to-one correspondance with
elements of $H^{n,0}(s\vert d)$, {\it i.e. }  solutions $a$ of the
equation 
\be s a+ d b =0\,, 
\label{topeq}
\ee 
with form-degree $n$
and ghost number zero, modulo the equivalence relation
$$a\sim a+sp+dq\,.$$

\noindent Quite generally, one can expand $a$ according to the antighost
number, as 
\be a=a_0+a_1+a_2+ \ldots a_k\,,
\label{antighdec}
\ee
where $a_i$ has antighost number $i$.  The expansion stops at some
finite value of the antighost number by locality, as was proved in
\cite{Barnich:1994mt}.

Let us recall \cite{Henneaux:1997bm} the meaning of the various
components of $a$ in this expansion. The antifield-independent
piece $a_0$ is the deformation of the Lagrangian; $a_1$, which is
linear in the antifields $h^{*}$, contains the information
about the deformation of the gauge symmetries, given by the
coefficients of $h^{*}$; $a_2$ contains the information
about the deformation of the gauge algebra (the term $C^{*} C C$
gives the deformation of the structure functions appearing in the
commutator of two gauge transformations, while the term $h^* h^* C
C$ gives the on-shell closure terms); and the $a_k$ ($k>2$) give
the informations about the deformation of the higher order
structure functions and the reducibility conditions.

\subsection{Equations}

In fact, using the previous cohomological theorems and standard
reasonings (see e.g. \cite{Boulanger:2000rq}), one can remove all components of $a$ with antifield
number greater than 2. The key point is that the invariant characteristic
cohomology $H^{n,inv}_k(\delta \vert d)$ controls the obstructions to
the removal of the term $a_k$ from $a$ and that all 
$H^{n,inv}_k(\delta \vert d)$ vanish for $k>2$ by Proposition \ref{usefll} and 
Proposition \ref{invKo}. 

Let us now decompose the cocycle condition (\ref{topeq}) according to the antighost number. If $a=a_0+a_1+a_2$, then $b$ can be assumed to stop at antigh 1 thanks to Proposition \ref{csq}. Using the fact that $s=\d+\g$, we obtain: 
\begin{eqnarray}
\g a_2=0 \,,\label{eqa2}\\ 
\d a_2 + \g a_1 + d b_1=0 \,,\label{eqa1} \\ 
\d a_1+\g a_0+d b_0=0 \,.\label{eqa0} 
\end{eqnarray} 
The first equation clearly means that $[a_2]\in H^2(\g) \Leftrightarrow a_2=\a_J \omega^J+\g c_2$ as in Equation (\ref{alom}). 
Applying $\g$ to Equation (\ref{eqa1}), $d\g b_1=0$ is obtained. 
Thanks to the Poincar\'e lemma and Proposition \ref{csq}, we see that $b_1$ can be taken in $H^2(\g)$ too : $b_1=\b_J \omega^J$. The second equation becomes $$(\d \a_J)\o^J+\g a_1+d\b_J \o^J+\b_J d\o^J=0\,.$$ 
Let us now introduce the differential $D$ defined in Section \ref{Ddeg}, we obtain 
$$(\d \a_J+d \b_J+\b_I A^I_J)\o^J=\g(...)=0\,.$$ 
This is because the left-hand side is strictly non $\g$-exact. Let us label the ghosts more precisely $\o^{J_i}$ where $i$ is the $D$-degree. Then, as $D$ raises by 1 the $D$-degree, the only non zero components of the matrix $A$ are $A^{J_i}_{J_{i+1}}$ and the last equation decomposes into: 
\begin{eqnarray}
\begin{array}{c}\d\a_{J_0}+ d\b_{J_0}=0 \,,\\ 
\forall i>0\ : \d \a_{J_i}+ d\b_{J_i}+ \b_{J_{i-1}}A^{J_{i-1}}_{J_i}=0\,.
\end{array}
\nonumber
\end{eqnarray}

\noindent The first equation means that $$\a_{J_0}\in H_2^n(\d|d,H_0(\g))\Rightarrow \a_{J_0}=[\l_{J_0 a \m}C^{*a\m}+\l_{J_0A\m\n}C^{*A\m\n}]d^n x\,,$$ 
thanks to Proposition \ref{H2} and Proposition \ref{invKo}. 
The $\l$'s are constants, because of the Poincar\'e invariance. 
We obtain $$\b_{J_0}=\frac{-1}{(n-1)!}\,[2\l_{J_0 a\m}h^{*a\m\a}+3\l_{J_0 A \m\n}C^{*A\m\n\a}]
\varepsilon_{\a\m_1\ldots\m_{n-1}}dx^{\m_1}\ldots dx^{\m_{n-1}}\,.$$ 

Thus, $\b_{J_0}A^{J_0}_{J_1}$ depends only on the underivated antifields, which cannot be 
$\d$-exact modulo $d$ unless they vanish, because a $\d$-exact term depends on the derivatives 
of the antifields or on the equations of motion, and because of the Poincar\'e invariance. 
Thus, $\d \a_{J_1}+d\b_{J_1}=0$ and $\b_{J_0}A^{J_0}_{J_1}=0$ independently. By applying the 
same reasoning recursively, the same decomposition appears to occur at every $D$-degree, so we 
finally obtain: 
\begin{eqnarray}\label{thisrocks}\begin{array}{c}
	\forall i\ : \d \a_{J_i} + d \b_{J_{i-1}} = 0
	\Rightarrow \a_{J_i} \in H_{2,inv}(\d|d)\,,\\ 
	\forall i\ : \b_{J_i}A^{J_i}_{J_{i+1}}=0\,.
\end{array}
\end{eqnarray}

%---------------------------------------
\subsection{Classification of the gauge algebras}
%---------------------------------------

The first set of equations (\ref{thisrocks}) provides a very limited number of candidates 
$a_2\,$. The different possible Lorentz-invariant terms have the form $\a_{J_i}\o^{J_i}=\l_{{J_i}...}\, C^{*...}\,\o^{J_i}\,$. 
The indices $J_i$ are spacetime indices and internal indices\footnote{For example: $\{\o^{J_0}\}=\{C^{a\a}C^{b\b},C^{a\a}C^{b\m\n},C^{A\a\b}C^{B\m\n}\}$.}.  
The Poincar\'e and parity invariance requirements impose that the constant tensors 
$\l_{{J_i}...}$ depend only on $\eta_{\m\n}$ or $\d^\a_\b$. Thus, the only possible terms are given by the Lorentz-invariant contractions of an undifferentiated antifield $C^{*...}$ with 
$\o^{{J_i}}$'s quadratic in the ghosts, contracted with arbitrary internal constant ``tensors''.
The pure spin-2 and spin-3 terms have already been studied in \cite{Boulanger:2000rq} and \cite{Bekaert:2005jf}. Let us give the exhaustive list of cross-interacting terms : 
\begin{eqnarray}
\stackrel{(1)}{a}_2 &=& \stackrel{(1)}{f}_{A[bc]}\,C^{*A\m\n}\,C_{\m}^b\,C_{\n}^c\,d^n x\,,
%\label{a21}
\nonumber
\\
\stackrel{(2)}{a}_2 &=& \stackrel{(2)}{f}_{aBc}\,C^{*a\m}\,C^B_{\m\n}\,C^{c\n}\,d^n x\,,
\nonumber
%\label{a22}
\\
\stackrel{(3)}{a}_2 &=& \stackrel{(3)}{f}_{ABc}\,C^{*A\m\n}\,\pa_{[\n}C^B_{\s]\m}\,
C^{c\s}\,d^n x\,,
%\label{a23}
\nonumber \\
\stackrel{(4)}{a}_2 &=& \stackrel{(4)}{f}_{ABc}\,C^{*A\m\n}\,C^{B\,\a}_{~\m}\,\pa_{[\n}C^c_{\a]}\,d^n x\,,
%\label{a24}
\nonumber \\
\stackrel{(5)}{a}_2 &=& \stackrel{(5)}{f}_{aBC}\,C^{*a\a}\,C^{B\m\n}\,
\pa_{[\n}C^C_{\a]\m}\,d^n x\,,
%\label{a25}
\nonumber \\
\stackrel{(6)}{a}_2 &=& \stackrel{(6)}{f}_{abC}\,C^{*a\m}\,\pa^{[\n}C^{\r] b}\,
\pa_{[\n}C^C_{\r]\m}\,d^n x\,,
%\label{a26}
\nonumber \\
\stackrel{(7)}{a}_2 &=& \stackrel{(7)}{f}_{[ab]C}\,C^{*C\m\n}\,\pa_{[\m}C_{\a]}^a \,\pa_{[\n}C_{\b]}^b\, \eta^{\a\b} \,d^n x\,,
%\label{a27}
\nonumber \\
\stackrel{(8)}{a}_2 &=& \stackrel{(8)}{f}_{aBC}\,C^{*a\m}\,\pa^{[\a}C^{|B|\b]\n}\,
\pa_{\m}\pa_{[\a}C_{\b]\n}^C\,d^n x\,,
%\label{a28}
\nonumber
\end{eqnarray}
where $\{\stackrel{(i)}{f}\}_{i=1}^8$ are eight arbitrary constant tensors.

Note that $a_2^{(8)}$ is trivial when $n=3\,$, because of a Schouten identity (or equivalently because there is no non-vanishing tensor $\widehat{U}_{\a\b|\m\n}$ in dimension 3).

%
%------------------------------------
\subsection{Computation of the gauge transformations}
%----------------------------------
%

The second set of equations (\ref{thisrocks}) has to be satisfied in order for Equation (\ref{eqa1}) to have a solution, and thus in order for $a_1$ to exist. This is not true for every $\stackrel{(i)}{a}_2$. In fact, it is faster to directly compute $\d a_2$ and check whether it is $\g$-exact modulo $d$, possibly given a symmetry rule on the internal indices. The latter condition $\d a_2 + \g a_1 + dc_1 = 0$ implies that the constant tensors $\stackrel{(1)}{f}_{Abc}$, $\stackrel{(2)}{f}_{aBc}$ and $\stackrel{(5)}{f}_{aBC}$ must vanish. The following relations between constant tensors are also obtained: 
 $\stackrel{(8)}{f}_{aBC}=\stackrel{(8)}{f}_{a(BC)}$ and $\stackrel{(4)}{f}_{ABc}=-\frac{3}{2}\stackrel{(3)}{f}_{ABc}$\,.
We thus get all of the possible $a_1$'s, which we classify 
according to the number of derivatives they involve: 
\begin{eqnarray}
	a_{1,1} &=&\stackrel{(3)}{f}_{ABc} \left[ \frac{3}{2}h^{*A\m\n\r}\left( \6_{[\n}h^B_{\s]\m\r}C^{c\s}-\6_{[\n}C^B_{\s]\m}h_\r^{c\s}-h_{\m\r}^{A\ \a}\6_{[\n}C^c_{\a]}+3C_\m^{B\a}\6_{[\n}h^c_{\a]\r}\right)\right.
	\nonumber 
\\ 
&& \left. \qquad\quad -\frac{3}{4n}h^{*\r}\6_\r(h_\s^B C^{c\s})\right]d^n x\,,
	\label{a11}
\\ 
\nonumber
		a_{1,2} &=& \stackrel{(6)}{f}_{abC}h^{*a\m\s}\left[2\6^\n h^{b\r}_\s \6_{[\n}C^C_{\r]\m}-\6^\n C^{b\r}\6_{[\n}h^C_{\r]\m\s}\right]d^n x 
\\ 
&& + 6 \stackrel{(7)}{f}_{abC}\left[h^{*C\m\n\r}-\frac{1}{n}\eta^{\m\n}h^{*C\r}\right]
\6_{[\m}h^a_{\a]\r}\6_{[\n}C^b_{\b]}\eta^{\a\b} d^n x + \bar{a}_{1,2}\,,
	\label{a12}
\\
	a_{1,3} &=&-\stackrel{(8)}{f}_{aBC}h^{*a\m\r}\6^\a C^{B\b\n}[2\6_{\m[\a}h^C_{\b]\n\r}-\6_{\n[\a}h^C_{\b]\m\r}]d^n x+\bar{a}_{1,3}\,,
	\label{a13} 
\end{eqnarray}
where the $\bar{a}_{1,i}$ terms are solutions of the homogeneous equations
\begin{eqnarray}
	\gamma \bar{a}_{1,i} + dc_{1,i} = 0\,,
\label{ab1}
\end{eqnarray}
which is equivalent to solving the equation 
\begin{eqnarray}
	\gamma \bar{a}_{1,i}= 0\,\Rightarrow [\bar{a}_{1,i}]\in H^1(\g)\,,
\label{abar1}
\end{eqnarray}
as proved by Proposition \ref{csq}. 

As a matter of fact, the solutions $\bar{a}_{1,i}$ linear in the fields play a crucial role 
in the present analysis, as we show in the next subsection. There is no such solution with one derivative, because the $\g$-closed functions of the fields involve at least 2 derivatives.

\subsection{Computation of cubic vertices}
%-----------------------------------------------

We now have to solve equation (\ref{eqa0}) for each $a_{1,i}\,$. 
In order to achieve this heavy calculation, we have been using FORM, a powerful software for 
symbolic computation (see \cite{form}). We have simply considered the most general candidates 
for $a_{0,i}$, implemented the $a_{1,i}$ (including the most general expression for 
$\bar{a}_{1,i}$) then solved the systems. It turns out that $\d a_{1,1}$ cannot be $\g$-exact 
modulo $d$, but we obtain consistent vertices for the two other cases, corresponding to 
$a_{1,2}$ and $a_{1,3}\,$. 

Incidentally, note that the last term on the first line of Equation (\ref{a11}) 
gives the first-order correction needed in order to transform the ordinary derivative $\partial_{(\m}\l_{\n\r)}$ into the covariant one $\nabla_{(\m}\l_{\n\r)}$ for the 
torsionless metric connection $\nabla$ associated with $g_{\m\n}=\eta_{\m\n}+h_{\m\n}\,$  
(where we omit to write the internal indices). In other words, we recover the result of
\cite{AL} that both minimal and non-minimal couplings of massless spin-3 field to 
dynamical gravity in flat space are inconsistent. In fact our result is more general.
It says that, whatever the complete gauge transformation of the spin-3 field is, 
if it contains the first-order correction needed to transform  
${\delta}_0 h_{\m\n\r}=3\,\partial_{(\m}\l_{\n\r)}$ into the covariantized transformation ${\delta}_{\l} h_{\m\n\r}=3\,\nabla_{(\m}\l_{\n\r)}$, then 
there is no consistent couplings between the spin-3 and spin-2 fields in flat background.  
Apart from Lorentz invariance, we made no assumptions on the form of the 
possible interaction and imposed no constraints on the number of derivatives (except that
it should be finite for locality). The result follows from consistency. 

%..............................................
\subsubsection{Solution with three derivatives}
\label{suba7}
 
First, some new relations on the structure constants are obtained: $\stackrel{(7)}{f}_{[ab]C}=-\frac{1}{3}\stackrel{(6)}{f}_{abC}$. 
{}From now on, we will call the structure constant $g_{Abc}=g_{A[bc]}$. The solution of Equations (\ref{eqa2}), (\ref{eqa1}) and (\ref{eqa0}) is as follows : 

\begin{eqnarray}a_{2,2}= g_{Abc}\left[\,C^{*b\m}\,\pa^{\n}C^{c\r}\,\pa_{[\n}C^A_{\r]\m}\,
 -\frac{1}{3}\,C^{*A\m\n}\,\pa_{[\m}C_{\a]}^b \,\pa_{[\n}C_{\b]}^c\,\eta^{\a\b}\right]
  \,d^n x\,,
\label{a72} 
\end{eqnarray}
\begin{eqnarray}a_{1,2} &=& g_{Abc}h^{*b\m\s}\left[2\6^\n h^{c\r}_\s 
\6_{[\n}C^A_{\r]\m}-\6^\n C^{c\r}\6_{[\n}h^A_{\r]\m\s}\right]d^n x
\nonumber \\ 
&& 
 - 3 g_{Abc}h^{*b\a\b}K^c_{\a\m\b\n}C^{A\m\n}d^n x 
\nonumber\\ 
&& 
-2 g_{Abc}\left[h^{*A\m\n\r}-\frac{1}{n}\eta^{\m\n}h^{*A\r}\right]
\6_{[\m}h^b_{\a]\r}\6_{[\n}C^c_{\b]}\eta^{\a\b} d^n x\,,
\end{eqnarray}
The expression for $a_{0,2}$ is given in Appendix \ref{vertex}. 

%.............................................
\subsubsection{Solution with four derivatives}
\label{suba8}

Let us rename $f_{aBC}$ the structure constant $\stackrel{(8)}{f}_{aBC}$. 
The solution of the equations (\ref{eqa2}), (\ref{eqa1}) and (\ref{eqa0}) is 
\begin{eqnarray} 
a_{2,3} = f_{aBC}\,C^{*a\m}\,\pa^{\a}C^{B\b\n}\,\,
\pa_{\m}\pa_{[\a}C_{\b]\n}^C\,d^n x\,,
\label{a82}
\end{eqnarray}
\begin{eqnarray} a_{1,3} &=& 
f_{aBC}\left[\frac{3}{8}h^{*a\m\n}\6^\r F^B_\r C^C_{\m\n}+\frac{3}{2}h^{*B\a\b\g}\6_\a 
K^a_{\b\m\g\n}C^{C\m\n}+\frac{2}{n}h^{*B\a}K_{\a\m\n\r}^a \6^\n C^{C\r\m}\right]d^n x
\nonumber\\ &&- 
\;f_{aBC}\,h^{*a\m\r}\,\6^\a 
C^{B\b\n}\left[2\6_{\m[\a}h^C_{\b]\n\r}-\6_{\n[\a}h^C_{\b]\m\r}\right]d^n x\,,
\label{a81}
\end{eqnarray}
The expression for $a_{0,3}$ is given in Appendix \ref{vertex}.

%*********************************
\section{Further results}
\label{sec:further}
%*********************************

%-----------------------------------------------
\subsection{Conditions at second order in the coupling constants}
%-----------------------------------------------

After this exhaustive determination of the consistent first-order deformations $W_1$, it is natural to study the second order equation (\ref{secor}). 
This equation can be decomposed into equations of definite antighost numbers. 
Let us consider $W_2=\int (b_0+b_1+b_2+...)$. The top equation is then $$(a_2,a_2)=-2\d b_3 -2\g b_2 + d(...)\,,$$ 
but $(a_2,a_2)$ does not depend on the fields and on the $antigh$-1 antifields so no $\d$-exact term can appear. This means that $\forall i>2\ :\ s b_i=d(...)$. 
This follows the same pattern as for $W_1$. 
It has been shown in \cite{Boulanger:2000rq} that the homology of $s$ modulo $d$ is trivial 
in that sector and thus the expansion of $W_2$ stops at $antigh = 2$.

Let us now compute the antibracket $(a_2,a_2)$. We have to consider for $a_2$ the sum of the different terms related to cubic vertices involving spin-2 and/or spin-3 fields. This includes $a_{2,2}$ and $a_{2,3}$, the pure spin-2 Einstein--Hilbert term \cite{Boulanger:2000rq} $$a_{2}^{EH}=a_{abc}C^{*a\mu}C^{b\nu}\6_{[\mu}C^c_{\nu]}d^n x$$ 
and the two pure spin-3 terms \cite{Bekaert:2005jf}
$$a_{2}^{BBvD}=k^A_{\ BC}\,C^{*\m\n}_{A} (T^B_{\m\a|\b}T^{C\a|\b}_{\n} 
-2T^B_{\m\a|\b}T^{C\b|\a}_{\n} + \frac{3}{2}\,C^{B\,\a\b}U^C_{\m\a|\n\b})\, d^nx$$ 
and 
$$a_{2}^{BBC}=l^{A}_{\ BC}\,C^{*\m\n}_{A}U^B_{\m\a|\b\l}U^{C\a|\b\l}_\n\,d^nx\,.$$ 
Let us give the list of the different antibrackets involving $a_{2,2}$ and $a_{2,3}$ in which we have already isolated $\g$-exact and $d$-exact parts : 
\begin{itemize}
\item[] $\begin{array}{rcl}(a_{2}^{EH},a_{2,2})&=&
-a^a_{~cd}g_{abE}C^{*c\n}\6_{[\n}C^d_{\m]}\6_{[\s}C^b_{\r]}T^{E\s\r|\m}
+a^a_{cd}g_{abE}C^{*c}_\t C^d_\m \6_{[\s}C^b_{\r]}U^{E\t\m|\s\r}\nonumber 
\\&&+ a^a_{\ cd}g_{ab}^{\ \ E}\6_{[\t}C^c_{\b]}\6_{[\m}C^d_{\n]}\eta^{\n\b}
\left(C^{*b\a}\6^{[\t}C^{\m]}_{E\a}+\frac{2}{3}C^{*\g[\t}_E\eta^{\m]\s}\6_{[\g}C^b_{\s]}\right)\nonumber\\&&+\g(...)+div.\end{array}$
\item[] $\begin{array}{rcl}(a_{2}^{EH},a_{2,3})&=&-a^a_{\ ef}f_{aBC}C^{*e\n}
\6_{[\n}C^f_{\m]}T^B_{\r\s|\t}U^{C\r\s|\m\t}
+a^a_{\ ef}f_{aBC}C^{*e\t}C^f_\m U^B_{\r\s|\t\n}U^{C\r\s|\m\n}\nonumber\\&&
+\g(...)+div.\end{array}$
\item[] $\begin{array}{rcl}(a_{2,2},a_{2,2})&=&
2g_{aeC}g^e_{\ bD}\left[C^{*a\m}\6_{[\b} C^b_{\g]} T^{C}_{\t\a|\m}U^{D\t\a|\b\g}
-\frac{2}{3}C^{*c\m\t}\6_{[\m}C^a_{\a]}\6_{[\b} C^b_{\g]}U_\t^{C\a|\b\g}\right]
\nonumber\\&&
+\g(...)+div.\end{array}$
\item[] $\begin{array}{rcl}(a_{2,2},a_{2,3})&=&g_{abC}f^a_{~DE}\left[C^{*b\n}T^C_{\t\m|\n}
+\frac{2}{3}C^{*C\n}_{\ [\t}\d^\s_{\m]}\6_{[\n}C^b_{\s]}\right]
U^{D\b\t|\r\a}U_{\b\ \ |\r\a}^{E\m}+\g(...)+div.\end{array}$
\item[] $\begin{array}{rcl}(a_{2,2},a_{2}^{BBvD})&=&g_{abC}k^C_{\ DE}C^{*a\m}\6_{[\t}C^b_{\n]}\times\\&&\left[\frac{3}{4}U^{D\t\n|\a\b}T^E_{\a\b|\m}
-3U_{\m(\a|\b)}^{D\quad\ \n}T^{E\t(\a|\b)}
+\frac{3}{4}U^{D\n}_{\m\ \ |\a\b}T^{E\a\b|\t}\right]
\nonumber\\ &&
+\frac{1}{2}g_{abC}k^C_{~DE}C^{*D}_{\a\b}
U^{E\a\m|\b\n}\6_{[\m}C^a_{\r]}\6_{[\n}C^b_{\s]}\eta^{\r\s}+\g(...)+div.\end{array}$
\item[] $\begin{array}{rcl}(a_{2,3},a_{2,3})&=&0\end{array}$
\item[] $\begin{array}{rcl}(a_{2,3},a_{2}^{BBvB})&=&-f_{aBC}k^B_{\ DE}
C^{*a}_\t U^{C\r\m|\t\n}\left[\frac{9}{8}U^D_{\r\m|\a\b}T^{E\a\b}_{\quad\ \ |\n}
+3U^D_{\r(\a|\b)\n}T^{D(\a|\b)}_\m\right]
\nonumber\\&&
+f_{aBC}k^B_{\ DE}C^{*a\r}T^{C\s\m|\n}\left[\frac{9}{8}U^D_{\r\n|\a\b}
U^{E~\,\a\b}_{\s\m |}+3U^D_{\r(\a|\b)\s}U_{\m\qquad\n}^{E(\a|\b)}\right]
\nonumber\\&&
+\g(...)+div.\end{array}$
\end{itemize}

\noindent The antibrackets involving $a_{2}^{BBC}$ are both $\g$-exact modulo $d$, which is mostly because of its high number of derivatives. We can see that $a_{2,3}$ seems to be behaving well. First, its antibracket with itself is vanishing. Then, looking at its antibracket with $a_{2}^{BBvD}$, it appears that the two expressions between square 
brackets of non $\g$-exact terms are very similar. 
The only difference is the ordering of the internal indices and thus, by imposing the relation 
$f_{aB(C}k^B_{D)E}=0$, this antibracket is consistent. Then, we can see that $a_{2,2}$ seems to 
be obstructed. The antibrackets $(a_{2,2},a_{2,2})$ and $(a_{2,2},a_{2}^{BBvD})$ have some non 
$\g$-exact terms in common but there is no combination allowing to remove them all. The only 
remaining ways to remove such terms would be to consider $a_2$ terms involving combinations of 
spins $1-2-2$, $1-3-3$, $1-2-3$ or $2-3-4$. As we stated in the introduction, it is already 
known that there is no deformation of the algebra for combinations $s_1-s_2-s_2$ where 
$s_1\geqslant 2s_2$ so there is no $a_2$ for a $4-2-2$ combination. It is very easy to see that 
there is no Lorentz-invariant way to write an $a_2$ for a $1-2-3$ combination. The remaining 
cases are more promising and we intend to study them in a near future. But it already appears 
to us that they do not remove the obstruction in $(a_{2,2},a_{2,2})\,$. 
It is not very surprising that 
the antibrackets $(a_{2}^{EH},a_{2,2})$ and $(a_{2,2},a_{2,3})$ do not behave well either. 
The last term, $(a_{2}^{EH},a_{2,3})$, brings in another interesting feature. 
Once again, the obstructions that it brings could be eliminated, 
resorting to $2-3-4$ interactions. However, as a preliminary computation shows, 
no terms quadratic in $U$ appear for $2-3-4$ deformations. 
So this indicates that the $2-3-3$ deformation that we 
have found is not compatible with the Einstein--Hilbert deformation. 

%---------------------------------------
\subsection{Gauging of rigid symmetries}
%---------------------------------------

In the previous section, we obtained two classes of first-order deformations
associated with non-Abelian algebras. Corresponding to the gauge-algebra deformation
$a_{2,2}$ displayed in equation (\ref{a72}) we have the following first-order 
gauge transformations
\begin{eqnarray}
 \delta h_{\a\b}^a &=& g^{~\,a}_{B~b} \left( 3 K^b_{\m\a \b\n}\l^{B\m\n} + 
	\partial_{[\n}h^b_{\rho]\a} \partial^{\n}\l^{B\rho}_{~~\b}
	+ \partial_{[\n}h^b_{\rho]\b} \partial^{\n}\l^{B\rho}_{~~\a} 
	- \partial_{[\n}h^B_{\rho]\a\b} \partial^{\n}\epsilon^{b\rho}
	\right)\,,
\label{tran2a} \\
  \delta h^A_{\m\n\rho} &=& 
  -2 g^A_{~bc} \left( \partial_{[\m}h_{\s]\rho}^b \partial_{[\n}\epsilon^c_{\t]}\eta^{\s\t}
  -\frac{1}{n}\,\eta_{\m\n} \,\partial_{[\a}h^b_{\b]\rho}\partial^{[\a}\epsilon^{\b]c}
  \right)\,,
  \label{tran3a}
\end{eqnarray}
where the right-hand-side of (\ref{tran3a}) must be totally symmetrized on the free indices. 
The first term on the right-hand-side of (\ref{tran2a}) comes from $\bar{a}_{1,2}$,  
solution of the equation (\ref{abar1}). The term 
$\bar{\delta}_{\lambda}h_{\a\b}^a=-3\,g^a_{~bB}\,K^b_{\m\a \n\b}\l^{B\m\n}$
is therefore absolutely necessary for the existence of the cubic vertex associated with
$a_{2,2}\,$. Up to some trivial gauge transformation, 
it is possible to express $\bar{\delta}_{\lambda}h_{\a\b}^a$ differently: 
	$\bar{\delta}_{\lambda}h_{\a\b}^a = \frac{3}{2}\,\cs_{\!\!_{\lambda}}\,h^a_{\a\b} 
	-3\,\partial_{(\a}V^a_{\b)}$,  
where 
\begin{eqnarray}
	\cs_{\!\!_{\lambda}}\,h^a_{\a\b} = 
	g^a_{~bB}\,\left[\lambda^{B\m\n}\,\partial_{\m}\partial_{\n}h^b_{\a\b}+
	\partial_{\a}\lambda^{B\m\n}\,\partial_{\m}h^b_{\n\b}+
	\partial_{\b}\lambda^{B\m\n}\,\partial_{\m}h^b_{\n\a}+
	\partial_{\a}\partial_{\b}\lambda^{B\m\n}\,h^b_{\m\n}\right]
	\label{Sder}
\end{eqnarray}
and 
$V^a_{\a}=g^a_{~bB}\,[2\,\lambda^{B\m\n}\,\partial_{[\m}h^b_{\a]\n}+\frac{1}{2}\,\partial_{\a}
(\lambda^{B\m\n}h^b_{\m\n})]\,$. The transformations 
$\frac{3}{2}\, \cs_{\!\!_{\lambda}}\,h^a_{\a\b}$
and $\bar{\delta}_{\lambda}h_{\a\b}^a $ are identified since they differ by a trivial 
zeroth-order gauge transformation $-3\,\partial_{(\a}V^a_{\b)}$ which can be eliminated 
by a redefinition of the gauge parameters $\epsilon_{\a}^a$ in Formula (\ref{freegauge}). 
Because the transformation $\bar{\delta}_{\lambda}h_{\a\b}^a$ involves the spin-2
fields $h_{\a\b}^a$ only \textit{via} the linearized Riemann tensors $K^b_{\m\a \n\b}$,
it is clear that $\bar{\delta}_{\lambda}h_{\a\b}^a $ is 
$\gamma$-closed and the following commutation 
relations hold $[\delta_{\epsilon},\bar{\delta}_{\lambda}]=0\,$, where 
$\delta_{\epsilon}h^a_{\a\b}=2\partial_{(\a}\epsilon^a_{\b)}\,$. 
The latter vanishing of commutator simply re-expresses the fact that $\bar{a}_{1}$ 
is not derived from any algebra-deformation $a_2$, since it satisfies the equation 
(\ref{ab1}) $\gamma \bar{a}_{1} + \delta a_2 = dc_1$ with $a_2$ $\d$-trivial modulo $d$.   

When the parameters $\lambda^{A\m\n}$ are all constant, the transformations (\ref{tran2a}) 
reduce, modulo trivial gauge transformations, to $\delta h^a_{\m\n} = \frac{3}{2}\,g^a_{~bB}\,\lambda^{B\a\b}\,\partial_{\a}\partial_{\b}h^b_{\m\n}$. 
Therefore, we recover the rigid symmetries 
$\delta h^a_{\m_1\ldots\m_s} \rightarrow h^a_{\m_1\ldots\m_s}+ \xi^{a~\a_1\ldots\a_n}_{~c}\,\partial_{\a_1}\ldots\partial_{\a_n}h^c_{\m_1\ldots\m_s}$
exhibited in \cite{Berends:1985xx}, where $s=2$, $n=2$ and 
$\xi^{ac}_{~~\a_1\a_2}=\frac{3}{2}\,\lambda^{B}_{\a_1\a_2}\,\delta^{cb}\,g^{~\;a}_{B\;b}\,$.    
As in \cite{Berends:1985xx}, we have $\xi^{ac}_{~~\a_1\a_2}=(-)^{n-1}\xi^{ca}_{~~\a_1\a_2}$
due to the symmetry properties of $g_{A\,bc}$ that we derived in our cohomological
analysis in Section \ref{suba7}. 
Retrospectively, we can therefore consider the consistent first-order deformation obtained in 
Section \ref{suba7} as resulting from the gauging of the rigid symmetry 
$h^a_{\m_1\m_2} \rightarrow  h^a_{\m_1\m_2}+ \xi^{a~\a_1\a_2}_{~c}\,\partial_{\a_1}\partial_{\a_2}h^c_{\m_1\m_2}$
presented in \cite{Berends:1985xx}. 
However, we have shown that this gauging is inconsistent when pushed
up to the second order in the coupling constants. 

\vspace*{.2cm}

\noindent\textbf{Remark:} 
In the framework of differential multicomplexes 
(see \cite{DVH} and the appendix A of \cite{Bekaert:2006ix} for 
precise definitions and related concepts) we are given
$s$ differential forms $d_ix^{\m}$ ($i=1,2,\ldots,s$) obeying 
$d_ix^{\m} d_jx^{\n}=(-)^{\delta_{ij}}d_jx^{\n} d_ix^{\m}\,$, where
wedge and symmetric products are not explicitly written.  
Their adjoints $(d_ix^{\m})^{\dagger}$ obey the same (anti)commutation 
relations and we have the crossed (anti)commutation relations  
$[d_ix^{\m},(d_jx^{\n})^{\dagger}]_{\pm}=\delta_{ij}\eta^{\m\n}$ where
$[~,~]_{\pm}$ stands for the graded commutator. 
The $s$ nilpotent differential operators 
$d_i=d_ix^{\m}\frac{\partial^L}{\partial x^{\m}}$ ($i=1,2,\ldots, s$) are 
defined which generalize the exterior differential of the de Rham complex.   
Setting $s=2$ and forgetting about the internal indices for a while since they
play no role in the following discussion, we may view $h_{\a\b}$ as the components
of the differential multiform $h=h_{\a\b}\,d_1x^{\a}d_2x^{\b}$ and re-write
Formula (\ref{Sder}) as
\begin{eqnarray}
	\cs_{\!\!_{\lambda}}\,h &=& (i_{\lambda}d_1d_2 + d_1i_{\lambda}d_2 + d_2i_{\lambda}d_1 + 
	d_1d_2i_{\lambda})\,h
  \nonumber \\
  &=& (d_1+d_2+i_{\lambda})^3\, h 
\label{LieDerS}
\end{eqnarray}
where
$i_{\lambda}=\lambda_{\m\n}(d_1x^{\m})^{\dagger}(d_2x^{\n})^{\dagger}\,$. 
The expression (\ref{LieDerS}) $\cs_{\!\!_{\lambda}}=(d_1+d_2+i_{\lambda})^3$ 
generalizes Cartan's formula 
$\cl_{\zeta}=(i_{\zeta}+d)(i_{\zeta}+d)=i_{\zeta}d+d\, i_{\zeta}$ for the usual 
Lie derivative along the vector $\zeta$ in the framework of the de Rham 
differential complex. There, $i_{\zeta}$ is the interior product associated 
with the vector $\zeta$ and $d$ the usual exterior differential. 

\noindent When a spin-1 field $A_{\m}$ is coupled to gravity, it transforms under 
diffeomorphisms via the Lie derivative along the diffeomorphism vector $\zeta$, 
$\delta^{diff}_{\zeta}A = \cl_{\zeta}A$ where $A=dx^{\m}\,A_{\m}\,$. 
By analogy, it is tempting to view $\cs_{\!\!_{\lambda}}\,h$ as the ``spin-3 diffeomorphism'' 
of the spin-2 field $h=h_{\a\b}\,d_1x^{\a}d_2x^{\b}$ 
along the gauge parameter $\lambda=\lambda^{\m\n}\partial_{\mu}\partial_{\nu}$ 
of the spin-3 field. However, as we said before, this transformation appears in 
the deformation $W_1=\int (a_{2,2}+a_{1,2}+a_{0,2})$ which, we have shown, 
is inconsistent when pushed at second order in the coupling constants. 
 
\vspace*{.2cm}
Let us now consider the gauge transformations corresponding to the 
four-derivative deformation $W_1=\int (a_{2,3}+a_{1,3}+a_{0,3})$
obtained in Section \ref{suba8}. See in particular Equation (\ref{a81}). 
It turns out to be convenient to add trivial gauge transformations
to them, by noting that 
\vspace{1mm}

\noindent $\quad\,\begin{array}{rcl}
\frac{3}{8}\,h^{*a\m\n}\6^\r F^B_\r C^C_{\m\n}&=&-\frac{3}{4n}\,
h^{*a\m\n}\delta(\partial^{\rho}h^{*B}_{\rho})C^C_{\m\n}=
\delta\left(\frac{3}{4n}\,
h^{*a\m\n}\partial^{\rho}h^{*B}_{\rho}C^C_{\m\n} \right)+\ldots
\nonumber \\
&=&
-\frac{3}{2n}\,h^{*B\rho}\left(\partial_{\rho}K^a_{\m\n}C^{C\m\n}+\frac{4}{3}\,K^{a\m\n}
\partial_{[\rho}C^C_{\m]\n}\right)+\delta(...)+\gamma(...)+d(...)\,	. 
\nonumber
\end{array}$

Accordingly, we have the following first-order gauge transformations of the 
spin-2 and spin-3 fields:
\begin{eqnarray}
\delta h^a_{\m\n} &=& -f^a_{~BC}\,\left[
\6^\a \lambda^{B\b\s}\left( 
\6_{\m[\a}h^C_{\b]\s\n}+\6_{\n[\a}h^C_{\b]\s\m}-\6_{\s[\a}h^C_{\b]\m\n}\right)
\right]
\label{transa8} \\
\delta h^B_{\a\b\c} &=& f^{~B}_{a~C}\left[\frac{3}{2}\,\left(
\6_\a K^a_{\b\m\g\n}-\frac{1}{n}\,\eta_{\a\b}\partial_{\gamma}K^a_{\m\n}
\right) \lambda^{C\m\n}\right.
\nonumber \\
&&\qquad\qquad\qquad
 \left. +\;\frac{2}{n}\,\eta _{\a\b}\left( K_{\c\m\n\r}^a + K^a_{\m[\n}\eta_{\r]\c}
\right)\6^\n \lambda^{C\r\m}\right]
\label{transb8}
\end{eqnarray}
where the right-hand-side of the second equation must be totally symmetrized over the free indices. 
By setting the gauge parameters to constants, 
one could wonder whether some rigid symmetry appears, that could retrospectively be seen
as being gauged. Clearly, nothing comes from the gauge transformations of $h^a_{\a\b}$. 
As far as the spin-3 fields are concerned, only the term  
$f^{~B}_{a~C}\left[\frac{3}{2}\,\left(
\6_\a K^a_{\b\m\g\n}-\frac{1}{n}\,\eta_{\a\b}\partial_{\gamma}K^a_{\m\n}
\right) \lambda^{C\m\n}\right]$ (which must be symmetrized over $\a\b\c$) 
survives when the gauge parameters are set to constants. 
However, with $\lambda^{C\m\n}$ all constants, it is readily seen that it can be 
written as $3\,\partial_{(\a}\Lambda^B_{\b\c)}$ where 
$\Lambda^B_{\b\c}=-f^{~B}_{a~C}\left[\frac{1}{2}\,\left(
K^a_{\m(\b\g)\n}+\frac{1}{n}\,\eta_{\b\c}K^a_{\m\n}\right) \lambda^{C\m\n}\right]$ and 
hence can be eliminated by a redefinition of the gauge parameters of the free
gauge transformations (\ref{freegauge3}). Indeed, $\eta^{\b\g}\Lambda^B_{\b\g}$ is 
vanishing. As a result, no rigid gauge transformations can be obtained 
upon setting to constants the gauge parameters in the
first-order gauge transformations (\ref{transa8}) and (\ref{transb8}).  
[Note, however, that the constants are not the only solutions of the higher-spin 
Killing equations.]

%********************************
\section{Conclusions and perspectives}
\label{sec:concl}
%********************************

In this paper we carefully analyzed the problem of introducing
consistent cross-interactions among a countable collection of
spin-3 and spin-2 gauge fields in flat spacetime of arbitrary dimension
$n> 3\,$. For this purpose we used the powerful BRST
cohomological deformation techniques in order to be exhaustive. 
Under the sole assumptions of locality, parity
invariance, Poincar\'e invariance and perturbative deformation
of the free theory, we proved that only two classes of
non-Abelian deformations are consistent at first order. 
The first deformation, which involves three derivatives in the Lagrangian,  
was already mentioned in the work \cite{Berends:1984rq}. We showed that
it is obstructed at second order in the deformation parameters if no 
other fields are present in the analysis.  
The second deformation involves four derivatives in the Lagrangian and passes
the second-order constraint --- equivalent to the Jacobi identities of the gauge
algebra at the corresponding order --- where the previous deformation failed. 
Moreover, combining this algebra-deformation with the one corresponding to
the Berends--Burgers--van Dam deformation, the crossed second-order constraint is also satisfied given a symmetry condition on the product of the internal coefficients, while a combination with the Einstein--Hilbert deformation is obstructed.

We also discussed the link between the gauge transformations associated with 
the three-derivative vertex and the rigid symmetries of the free theory 
exhibited in \cite{Berends:1985xx}. More precisely, these gauge transformations 
can be seen as a gauging of these rigid symmetries, similarly to what happens
in the ``Bell--Robinson" cases $s_1-s_2-s_2$ where $s_1\geqslant 2s_2$ 
\cite{Deser:1990bk,Berends:1985xx}.  

It would be of interest to enlarge the set of fields to spin $4$ 
and see if this allows to remove the previous obstructions at order two. 
A hint that this might be sufficient comes from the fact that the commutator of two 
spin-$3$ generators produces spin-$2$ and spin-$4$ generators for the bosonic 
higher-spin algebra of Ref. \cite{Vasiliev:2004qz}. We hope to address this issue
in the future. 

More generally, we believe that the two consistent vertices exhibited here can be related 
to the flat space limit (appropriately defined, in order to avoid potential problems related to the non-analyticity in the cosmological constant $\Lambda$) of the spin-3--spin-2 sector of the full $AdS_n$ higher-spin gauge theory of \cite{Vasiliev:2004qz,Sagnotti:2005,Fradkin:1987ks}
(and references therein).   
Such a connection would provide a geometric meaning for the long expressions for the 
vertices. That such a Minkowski--$(A)dS$ link should be possible was mentioned in the pure 
spin-3 case \cite{Bekaert:2005jf}. In fact, it can be shown that there is a  
correspondence between the non-Abelian gauge algebras obtained in the current flat-spacetime
setting and the $AdS_n$ higher-spin algebra $hu(1|2:[n-1,2])$ reviewed
e.g. in the second reference of \cite{Sagnotti:2005}. This is beyond the scope of 
the present work and will be reported elsewhere. 

\section*{Acknowledgments}
%*******************************************

We thank Xavier Bekaert, Augusto Sagnotti and Philippe Spindel for stimulating discussions.  
The work of N.B. is supported by the Fonds National de la Recherche Scientifique (Belgium). 
%\newpage
%*****************************************

\appendix
\section{Proof of Proposition 1}
%\section{Proof of Proposition \ref{Hgamma}}
\label{SergeA}

First, let us recall the results for pure spin-2 \cite{Boulanger:2000rq} and pure spin-3 
\cite{Bekaert:2005jf} theories. 
In the spin-2 case, a convenient set of representatives of the cohomology of $\gamma$ is 
the set of functions of the antighosts, $[K^a]$, $C^a_\mu$ and $\6^{}_{[\mu}C^a_{\n]}$. 
In the spin-3 
case, the natural set of representatives of the cohomology of $\gamma$ is the set of functions of the antighosts, $[K^A]$, $[F]$, $C^A_{\m\n}$, $\widehat{T}^A$ and $\widehat{U}^A$. 
Proposition \ref{Hgamma} just says that, in the mixed case, the cohomology of $\gamma$ is the direct product of the previous two sets.

\subsubsection*{$pgh\ 0$}

$\g$ can be seen as the sum of its spin 2 and spin 3 restrictions, that we will note $\g_2$ and $\g_3$. The homology of $\g_2$ (resp. $\g_3$) is simply the direct product of the homology in the spin 2 (resp. 3) case and the set of all spin 3 (resp. 2) fields.
Given an arbitrary $\g$-closed function $f$ at $pgh$ 0 (i.e. it does not depend on any ghost), we have $\g f = \g_2 f + \g_3 f = 0$.\\ But, as $\g_2 f$ is linear in the spin 2 ghosts and $\g_3 f$ is linear in the spin 3 ghosts, the two terms are linearly independant and thus both vanish. This means that $f$ is in the intersection of the homologies of $\g_2$ and $\g_3$, so $f=f([\Phi^*_I],[K^a],[K^A],[F^A])$. As there are no $\g$-exact objects in $pgh\ 0$, $H^0(\g)$ is the set of those functions.

\subsubsection*{$pgh>0$}

Let us denote generically $\cc^a$/$\cc^A$ a basis of the spin 2/spin 3 ghosts and their derivatives. The $\g$-exact ghosts will be noted $\bar{\cc}^a=\g [h]^a$ and $\bar{\cc}^A=\g [h]^A$ (the bracketed fields are the adequate combinations of the fields or some of their derivatives). A $pgh\ i$ object $f^i$ (with $antigh\ f^i=k$) is then a linear combination: $$f^i=\sum_j f_{a_1...a_jA_1...A_{i-j}}\cc^{a_1}...\cc^{a_j}\cc^{A_1}...\cc^{A_{i-j}}$$ Imposing that $\g f^i=0$ gives rise to some relations between the $\g f_{a_1...a_jA_1...A_{i-j}}$. In general, they would be a combination of the $\bar{\cc}^a$ and the $\bar{\cc}^A$, but the coefficients will have to take particular values: \begin{eqnarray}\exists\ \{\stackrel{(j)}{\ck}\}\ | \g f_{a_1...a_jA_1...A_{i-j}}&=&(-1)^k\ \stackrel{(j+1)}{\ck}_{\!\!\!a_1...a_{j+1}A_1...A_{i-j}}\bar{\cc}^{a_{j+1}}\nonumber\\&& + (-1)^{i+j+k}\stackrel{(j)}{\ck}_{a_1...a_jA_1...A_{i-j+1}}\bar{\cc}^{A_{i-j+1}}\nonumber\end{eqnarray} The coefficients $\ck$ must be taken such that at least one index $a$ and one index $A$ are contracted only with gamma exact objects (say the last index of both kinds as in the last equation). The $antigh$ sign factor has been introduced for later convenience. Finally, the $j=0$ and $j=i+1$ coefficients have to vanish.\\ Let us remark that $\g^2 f_{a_1...a_jA_1...A_{i-j}}=0$ implies that $\forall\ j\ :\ \g \stackrel{(j)}{\ck}=0$. This means that \begin{eqnarray}f_{a_1...a_jA_1...A_{i-j}}&=&\stackrel{(j+1)}{\ck}_{\!\!\!a_1...a_{j+1}A_1...A_{i-j}}[h]^{a_{j+1}}+ (-1)^{i+j}\stackrel{(j)}{\ck}_{a_1...a_jA_1...A_{i-j+1}}[h]^{A_{i-j+1}}\nonumber \\&&+ \,g_{a_1...a_jA_1...A_{i-j}}\nonumber \end{eqnarray} where $g_{a_1...a_jA_1...A_{i-j}}\in H^0(\g)$ and we obtain an expression for $f^i$ 
itself: 
\begin{eqnarray} f^i&=&\displaystyle\sum_j[ \stackrel{(j+1)}{\ck}_{\!\!\!a_1...a_{j+1}A_1...A_{i-j}}[h]^{a_{j+1}}\cc^{a_1}...\cc^{a_j}\cc^{A_1}...\bar{\cc}^{A_{i-j}} \nonumber\\&&\quad + (-1)^{i+j}\stackrel{(j)}{\ck}_{a_1...a_jA_1...A_{i-j+1}}[h]^{A_{i-j+1}}\cc^{a_1}...\bar{\cc}^{a_{j}}\cc^{A_1}...\cc^{A_{i-j}}\nonumber\\&&\quad +g_{a_1...a_jA_1...A_{i-j}}\cc^{a_1}...\cc^{a_j}\cc^{A_1}...\cc^{A_{i-j}}]\nonumber\\
&=&\displaystyle\sum_j[\stackrel{(j+1)}{\ck}_{\!\!\!a_1...a_{j+1}A_1...A_{i-j}}\cc^{a_1}...\cc^{a_j}\cc^{A_1}...\cc^{A_{i-j-1}}([h]^{a_{j+1}}\bar{\cc}^{A_{i-j}}+\bar{\cc}^{a_{j+1}}[h]^{A_{i-j}})\nonumber\\&&\quad+g_{a_1...a_jA_1...A_{i-j}}\cc^{a_1}...\cc^{a_j}\cc^{A_1}...\cc^{A_{i-j}}]\nonumber\\&=&\g\left((-1)^{i+k+1}\displaystyle\sum_j \stackrel{(j)}{\ck}_{a_1...a_{j+1}A_1...A_{i-j}}\cc^{a_1}...\cc^{a_j}\cc^{A_1}...\cc^{A_{i-j-1}}[h]^{a_{j+1}}[h]^{A_{i-j}}\right)
\nonumber \\
&&+\sum_j g_{a_1...a_jA_1...A_{i-j}}\cc^{a_1}...\cc^{a_j}\cc^{A_1}...\cc^{A_{i-j}}
\nonumber
\end{eqnarray}
The first term of the last expression is trivial in $H^i(\g)$ while the second, as we announced, depends on the fields only through $[K^a]$, $[K^A]$ and $[F^A]$ thanks to the fact that the coefficients $g$ belong to $H^0(\g)$.\\ Finally, the second term can be rewritten as \begin{eqnarray}\nonumber
g_{a_1...a_jA_1...A_{i-j}}\cc^{a_1}...\cc^{a_j}\cc^{A_1}...\cc^{A_{i-j}}&=&G_a\bar{\cc}^a+G_A\bar{\cc}^A+\alpha_I\omega^I 
\\ \nonumber
&=& \g\{G_a[h]^a+G_A[h]^A\}+\alpha_I\omega^I 
\end{eqnarray} where, as stated before, $\{\omega^J\}$ is a basis of the products of non-exact ghosts. The only non-exact term in the last equation is the last one, with $\a_J\in H^0(\g)$. This expression is the general form for a representative of $H^i(\g)$ that we announced.
%*******************************************

\section{First-order vertices}
\label{vertex}

The three-derivative first-order vertex corresponding to the algebra
deformation $a_{2,2}$ given in (\ref{a72}) is 
\begin{eqnarray}
  a_{0,2} = \stackrel{(3)}{\cal{L}}d^n x = g^A_{\ bc}\,U^{bc}_{A}\, d^n x \,,
\nonumber
\end{eqnarray}
where, denoting $h=  \eta^{\m\n}h_{\m\n}$ and $h_{\a}=  \eta^{\m\n}h_{\a\m\n}\,$, 
\begin{eqnarray}
	U^{bc}_{A} &=& -  \frac{1}{2}\, h_A^{\a} \Box  h^b \pa_{\a} h^c 
	        +  \frac{1}{2}\, h_A^{\a\b\c} \pa_{\b}\pa_{\c}h^b \pa_{\a}h^c
	        +  \frac{1}{2}\,h_A^{\a\b\c} \Box  h^b_{\b\c} \pa_{\a}h^c
\nonumber \\       
	   &&+\; \frac{1}{2}\,h_A^{\a} \pa^{\b}\pa^{\c} h^b_{\b\c} \pa_{\a}h^c
         +              h_A^{\a} \pa^{\b}\pa^{\c} h^b  \pa_{\a}h_{\b\c}^c
	       + \frac{1}{2}\,h_A^{\a} \Box  h^{b\b\c} \pa_{\a}h_{\b\c}^c
\nonumber \\       
	   && -\; h_A^{\a\b\c} \pa_{\b\d}h_{\c}^{b\,\d}\pa_{\a}h^c
	        - h_A^{\a\b\c} \pa_{\b\d}h^b \pa_{\a}h_{\c}^{c\,\d}
          - \frac{1}{2}\,h_A^{\a\b\c} \Box  h^b_{\b\d}\pa_{\a}h_{\c}^{c\,\d}
\nonumber \\       
	   &&-\; \frac{3}{2}\,h_A^{\a} \pa^{\b}\pa^{\c} h^b_{\b\d} \pa_{\a}h_{\c}^{c\,\d}
	        - \frac{1}{2}\,h_A^{\a\b\c} \pa_{\b}\pa_{\c} h^{b\m\n} \pa_{\a}h_{\m\n}^c
          - h_A^{\a\b\c} \pa^{\m}\pa^{\n} h^b_{\b\c} \pa_{\a}h_{\m\n}^c
\nonumber \\       
	   &&+\; 
	         \frac{1}{2}\,h_A^{\a\b\c} \pa_{\c}\pa_{\d} h^{b\d\e} \pa_{\a}h^c_{\b\e}
         + \frac{3}{2}\,h_A^{\a\b\c} \pa_{\c}\pa_{\d} h^b_{\b\e} \pa_{\a}h^{c\d\e}
	       + h_A^{\a\b\c} \pa^{\d}\pa^{\e} h^b_{\b\d} \pa_{\a}h^c_{\c\e}
\nonumber \\       
	   &&-\; \frac{1}{4}\,h_A^{\a} \pa_{\a}\pa_{\c}h^b  \pa^{\c}h^c
	        - \frac{1}{2}\,h_A^{\a\b\c} \pa_{\a}\pa_{\e}h^b_{\b\c}\pa^{\e}h^c
          + h_A^{\a\b\c} \pa_{\a}\pa_{\e}h^b \pa^{\e}h_{\b\c}^c
\nonumber \\       
	   &&+\; \frac{1}{4}\,h_A^{\a} \pa_{\a}\pa_{\e}h^{b\m\n} \pa^{\e}h_{\m\n}^c
	        - \frac{1}{2}\,h_A^{\a\b\c} \pa_{\a}\pa_{\e}h^b_{\b\d} \pa^{\e}h_{\c}^{c\,\d}
	        + h_{A\m} \pa_{\a}\pa_{\e}h^{b\a\m} \pa^{\e}h^c 
\nonumber \\       
	   &&-\; h_{A\m\b\c} \pa_{\a}\pa_{\e}h^{b\a\m} \pa^{\e}h^{c\b\c} 
	        + \frac{1}{2}\,h_A^{\t} \pa_{\a}\pa_{\e}h^{b\a\m} \pa^{\e}h^c_{\m\t} 
	        - h_{A\m} \pa_{\a}\pa_{\e}h^b \pa^{\e}h^{c\a\m} 
\nonumber \\       
	   &&+\; h_{A\m\b\c} \pa_{\a}\pa_{\e}h^{b\b\c} \pa^{\e}h^{c\a\m} 
          - \frac{1}{2}\,h_A^{\t} \pa_{\a}\pa_{\e}h^b_{\m\t} \pa^{\e}h^{c\a\m} 
	        - \frac{1}{2}\,h_A^{\a} \Box  h^b_{\a\c} \pa^{\c}h^c  
\nonumber \\       
	   &&+\; \frac{1}{2}\,h_A^{\a\b\c} \pa_{\b}\pa_{\c} h^b_{\a\r} \pa^{\r}h^c  
	        +\frac{1}{2}\,h_A^{\a\b\c} \Box  h^b_{\a\r} \pa^{\r}h_{\b\c}^c  
          +\frac{1}{2}\,h_A^{\a} \pa_{\b}\pa_{\c} h^b_{\a\r} \pa^{\r}h^{c\b\c}
\nonumber \\       
	   &&-\; h_A^{\a\b\c} \pa_{\b}\pa_{\d} h^b_{\a\r} \pa^{\r}h_{\c}^{c\,\d}  
	        -\frac{1}{4}\,h_{A\m} \pa^{\b}\pa^{\m} h^b_{\b\c} \pa^{\c}h^c  
          -\frac{1}{2}\,h_{A\m\n\r} \pa^{\b}\pa^{\m} h^b_{\b\c} \pa^{\c}h^{c\n\r}
\nonumber \\       
	   &&+\;   h_A^{\n} \pa^{\b}\pa^{\m} h^b_{\b\c} \pa^{\c}h_{\m\n}^c  
	        - \frac{1}{2}\,h_{A\m} \Box  h^b_{\b\c} \pa^{\c}h^{c\b\m}  
          + \frac{1}{2}\,h_{A\m\n\r} \pa^{\n}\pa^{\r} h^b_{\b\c} \pa^{\c}h^{c\b\m} 
\nonumber \\       
	   &&-\;\frac{1}{4}\,h_A^{\t} \pa_{\m}\pa_{\t} h^b_{\b\c} \pa^{\c}h^{c\b\m}  
	        + \frac{1}{2}\,h_A^{\a} \Box  h^b \pa^{\c}h_{\a\c}^c  
          - \frac{1}{2}\,h_A^{\a\m\n} \pa_{\m}\pa_{\n}h^b \pa^{\c}h_{\a\c}^c
\nonumber \\       
	   &&-\; \frac{1}{2}\,h_A^{\a\m\n} \Box  h^b_{\m\n} \pa^{\c}h_{\a\c}^c  
	        - \frac{1}{2}\,h_A^{\a} \pa^{\m}\pa^{\n} h^b_{\m\n} \pa^{\c}h^c_{\a\c}  
          + h_A^{\a\m\n} \pa_{\n}\pa_{\r} h_{\m}^{b\,\rho} \pa^{\c}h^c_{\a\c}
\nonumber \\       
	   &&+\; \frac{1}{4}\,h_{A\b} \pa^{\a}\pa^{\b} h^b \pa^{\c}h_{\a\c}^c  
	        + \frac{1}{2}\,h_{A\b\m\n} \pa^{\a}\pa^{\b} h^{b\m\n} \pa^{\c}h^c_{\a\c}  
          - h_A^{\t} \pa^{\a}\pa^{\m} h^b_{\m\t} \pa^{\c}h_{\a\c}^c
\nonumber \\       
	   &&+\; \frac{1}{2}\,h_{A\m} \Box  h^{b\a\m}\pa^{\c}h_{\a\c}^c
	        - \frac{1}{2}\,h_{A\m\n\r} \pa^{\n}\pa^{\r} h^{b\a\m}\pa^{\c}h^c_{\a\c}
          + \frac{1}{4}\,h_A^{\t} \pa_{\m}\pa_{\t} h^{b\a\m}\pa^{\c}h^c_{\a\c}  \,. 
\nonumber
\end{eqnarray}

%\vspace{.2cm}

\noindent The four-derivative first-order vertex corresponding to the algebra
deformation $a_{2,3}$ given in (\ref{a82}) is 
\begin{eqnarray} 
a_{0,3} = \stackrel{(3)}{\cal{L}}d^n x = f^a_{\ BC}T^{BC}_a d^n x\,, 
\nonumber
\end{eqnarray}
where
\begin{eqnarray}
\nonumber 
T^{BC}_a &=& \frac{1}{4}\, h_a \Box\pa^2_{\a\b}h_\g^B h^{C\a\b\g}
 - \frac{1}{4}\, h_a\pa^4_{\a\b\g\d}h^{B\a\b\ve}h^{C\g\d}_\ve 
 + \frac{1}{8}\, h_a\pa^4_{\a\b\g\d}h^{B\a} h^{C\b\g\d} 
\nonumber \\ 
&&-\;\frac{1}{4}\, h_a^{\a\b}\pa^4_{\a\b\m\n}h^B_\g h^{C\g\m\n} 
 - \frac{1}{2}\, h_a^{\a\b}\pa^4_{\a\m\n\r}h^B_\b h^{C\m\n\r} 
 + \frac{1}{2}\, h_a^{\a\b}\pa^4_{\a\m\n\r}h_\b^{B\r\s}h_\s^{C\m\n} 
\nonumber \\ 
&&+\;\frac{1}{4}\, h_a^{\a\b}\pa^4_{\a\m\n\r}h^{B\m}h_\b^{C\n\r}
 - \frac{1}{4}\, h_a^{\a\b}\Box\pa^2_{\m\n}h^B_{\a\b\r}h^{C\m\n\r}
 + \frac{1}{4}\, h_a^{\a\b}\pa^4_{\m\n\r\s}h^{B\m}_{\ \a\b}h^{C\n\r\s} 
\nonumber \\ 
&&+\; \frac{1}{2}\, h_a^{\a\b}\Box\pa^2_{\m\n}h^B_\a h_\b^{C\m\n} 
 - \frac{1}{2}\, h^{\a\b}_a\pa^{4\m\n\r\s}h^B_{\a\m\n}h^C_{\b\r\s} 
 - \frac{3}{8}\, h^{\a\b}_a\Box\pa^{2\m\n}h^B_\m h^C_{\a\b\n}
\nonumber \\ 
&&+\;  \frac{1}{4}\, h^{\a\b}_a\pa^{4\m\n\r\s}h^B_{\m\n\r}h^C_{\a\b\s}
\nonumber \\
&&-\; \frac{1}{2}\, h^{\a\b}_a\pa^3_{\b\m\n}h^B_\r\pa_\a h^{C\m\n\r} 
 + \frac{3}{8}\, h^{\a\b}_a\pa^3_{\b\m\n}h^{B\m}\pa_\a h^{C\n} 
 + \frac{1}{4}\, h^{\a\b}_a \pa^3_{\b\m\n} h^{B\m\r\s} \pa_{\a} h^{C\n}_{\ \ \r\s}
\nonumber \\ 
&&+\; \frac{3}{4}\, h^{\a\b}_a\Box\pa_{\g} h_\b^B\pa_\a h^{C\g}
 - \frac{1}{4}\, h^{\a\b}_a\Box\pa_{\g} h_{\b\m\n}^B\pa_\a h^{C\g\m\n}
 - \frac{1}{4}\, h^{\a\b}_a\pa^3_{\m\n\r}h_\b^B\pa_\a h^{C\m\n\r}
\nonumber \\ 
&&-\; \frac{3}{4}\, h^{\a\b}_a\pa^{3\m\n\r} h_{\b\n\r}^B\pa_\a h^{C}_\m 
 + \frac{1}{2}\, h^{\a\b}_a\pa^{3\m\n\r}h^B_{\b\m\s}\pa_\a h_{\n\r}^{C\ \s}
 - \frac{9}{8}\, h^{\a\b}_a\Box\pa_\g h^{B\g}\pa_\a h^C_\b
\nonumber \\ 
&&+\; \frac{3}{4}\, h^{\a\b}_a\pa^3_{\m\n\r} h^{B\m\n\r}\pa_\a h^C_\b 
 + \frac{1}{2}\, h_a\Box\pa^\a h^{B\m\n\r}\pa_\a h^C_{\m\n\r}
 + \frac{1}{2}\,h_a\pa^{3\a\m\n} h^{B\r}\pa_\a h^C_{\m\n\r}
\nonumber \\ 
&&+\; \frac{3}{8}\, h_a\pa^{3\a\m\n} h^B_\m \pa_\a h^C_\n 
 - h_a\pa^{3\a\m\n} h^B_{\m\r\s} \pa_\a h^{C\r\s}_\n 
 - \frac{1}{2}\, h_a^{\a\b}\pa^3_{\g\a\b} h^B_{\r\s\t} \pa^\g h^{C\r\s\t} 
\nonumber \\ 
&&-\; \frac{1}{2}\, h_a^{\a\b}\pa^3_{\g\m\n} h^B_{\a\b\r} \pa^\g h^{C\r\s\t}
 + \frac{3}{4}\, h_a^{\a\b}\pa^{3\g\m\n} h^B_{\a\b\m} \pa_\g h^C_\n 
 - \frac{1}{2}\, h_a^{\a\b}\pa^{3\g\m\n} h^B_{\m} \pa_\g h^C_{\a\b\n} 
\nonumber \\ 
&&-\; \frac{3}{2}\, h_a^{\a\b}\pa^3_{\g\a\m} h^B_\b \pa^\g h^{C\m}
 + h_a^{\a\b}\pa^3_{\g\a\m} h^B_{\b\n\r} \pa^\g h^{C\m\n\r}
 + \frac{3}{8}\, h_a^{\a\b}\pa^3_{\g\a\m} h^{B\m} \pa^\g h^C_\b 
\nonumber \\ 
&&+\; \frac{3}{4}\, h_a^{\a\b}\pa^3_{\g\a\m} h^{B\m\n\r} \pa^\g h^C_{\b\n\r} 
 + \frac{3}{4}\, h_a^{\a\b}\Box\pa_\g h^B_\a\pa^\g h^C_\b 
 - \frac{3}{4}\, h_a^{\a\b}\Box\pa_\g h^B_{\a\m\n}\pa^\g h^{C\m\n}_\b
\nonumber \\ 
&&-\; \frac{3}{4}\, h_a^{\a\b}\pa^{3\g\m\n} h^B_{\a\m\n}\pa_\g h^C_\b
 + \frac{3}{4}\, h_a^{\a\b}\pa^{3\g\m\n}h^B_\a\pa_\g h^C_{\b\m\n} 
 + \frac{3}{4}\, h_a\Box\pa^\a h^{B\b}\pa_\b h^C_\a 
\nonumber \\  
&&-\; \frac{1}{2}\, h_a\pa^{3\m\n\r} h^{B\a}\pa_\a h^C_{\m\n\r}
 - \frac{3}{4}\, h_a\pa^3_{\m\n\r}h^{B\a\m\n}\pa_\a h^{C\r} 
 - \frac{1}{2}\, h_a\Box\pa^\r h^{B\a\m\n}\pa_\a h^C_{\m\n\r} 
\nonumber \\ 
&&+\; h_a\pa^3_{\m\r\s}h^{B\a\m\n}\pa_\a h_\n^{C\r\s} 
 - \frac{3}{4}\, h_a^{\a\b}\Box\pa_\m h^B_{\g\a\b}\pa^\g h^{C\m}
 + \frac{1}{2}\, h_a^{\a\b}\pa^3_{\m\n\r} h^B_{\g\a\b}\pa^\g h^{C\m\n\r} 
\nonumber \\ 
&&-\; \frac{3}{4}\, h_a^{\a\b}\pa^3_{\m\a\b} h^B_\g\pa^\g h^{C\m} 
 + \frac{1}{2}\, h_a^{\a\b}\pa^3_{\m\a\b} h^B_{\g\n\r}\pa^\g h^{C\m\n\r}
 - h_a^{\a\b}\Box\pa^\m h^B_\g\pa^\g h^C_{\m\a\b}
\nonumber \\ 
&&+\; h_a^{\a\b}\pa^{3\m\n\r} h^B_{\g\n\r}\pa^\g h_{C\m\a\b} 
 - h^{\a\b}_a\pa^3_{\a\m\n}h^B_{\b\g\r}\pa^\g h^{C\m\n\r} 
 + \frac{3}{2}\, h^{\a\b}_a\pa^3_{\a\m\n}h_\b^{B\g\m}\pa_\g h^{C\n} 
\nonumber \\ 
&&-\; \frac{3}{2}\, h^{\a\b}_a\pa^{3\m\n\r}h^B_{\a\g\r}\pa^\g h^C_{\b\m\n} 
 + \frac{3}{2}\, h^{\a\b}_a\Box\pa^\m h_\a^{B\g\n}\pa_\g h^C_{\b\m\n} 
 + \frac{3}{2}\, h^{\a\b}_a\pa^3_{\a\m\n}h^B_\g\pa^\g h_\b^{C\m\n}
\nonumber \\  
&&-\; \frac{3}{2}\, h^{\a\b}_a\pa_{\a\m\n} h^{B\g\m\r}\pa_\g h_{\b\r}^{C\ \n} 
 + \frac{3}{4}\, h^{\a\b}_a\Box\pa_\m h^{B\m}\pa^\n h^C_{\n\a\b}
 - \frac{1}{2}\, h^{\a\b}_a\pa^3_{\m\n\r} h^{B\m\n\r}\pa^\s h^C_{\s\a\b} 
\nonumber \\  
&&-\; \frac{1}{4}\, h^{\a\b}_a \pa^3_{\a\m\n}h^{B\m}\pa_\g h_\b^{C\g\n}
 - \frac{1}{2}\, h^{\a\b}_a\Box\pa^\m h^B_\a\pa^\g h^C_{\b\g\m}
 + \frac{1}{2}\, h^{\a\b}_a\pa^{3\m\n\r} h^B_{\a\n\r}\pa^\g h^C_{\b\g\m}	
\nonumber \\ 
&&-\; \frac{1}{4}\, h_a\Box h^{B\a}\Box h^C_\a
 + \frac{1}{8}\, h_a\Box h^{B\a\b\g}\Box h^C_{\a\b\g}
 + \frac{1}{8}\, h_a \Box h^{B\a} \6^2_{\a\b} h^{C\b}
\nonumber \\
&&-\; \frac{3}{4}\, h_a \Box h^{B\r\s\a}\6^2_{\a\b} h^{C\ \b}_{\r\s}
 + \frac{3}{4}\, h_a\Box h^{B\a\b\m} \6^2_{\a\b} h^C_\m
 + \frac{1}{2}\, h_a\Box h^B_\m \6^2_{\a\b} h^{C\a\b\m}
\nonumber \\
&&+\; \frac{1}{4}\, h_a \6^{2\a\b}h^{B\m} \6^2_{\a\b}h^C_{\m}
 + \frac{3}{8}\, h_a \6^{2\a\b}h^{B\m\n\r} \6^2_{\a\b}h^C_{\m\n\r}
 + \frac{1}{8}\, h_a \6^{2\a\m}h^B_\m \6^2_{\a\n}h^{C\n}
\nonumber \\
&&-\; \frac{1}{8}\, h_a \6^{2\a\m}h^B_{\m\r\s} \6^2_{\a\n}h^{C\n\r\s}
 + \frac{7}{8}\, h_a \6^{2\a\m}h^{B\n} \6^2_{\a\n}h^C_{\m}
 - \frac{5}{8}\, h_a \6^{2\a\m}h^{B\n\r\s} \6^2_{\a\n}h^C_{\m\r\s}
\nonumber \\
&&-\; h_a \6^{2\a\m}h^B_{\m\n\r}\6_\a^{2\n}h^{C\r}
 - \frac{1}{4}\, h_a \6^{2\a\b}h^B_{\a\b\g} \6^2_{\m\n}h^{C\m\n\g}
 - \frac{1}{8}\, h_a \6^{2\a\b}h^B_{\a\b\g} \6^{2\g\m}h^C_\m
\nonumber \\
&&+\; \frac{1}{4}\, h_a \6^{2\a\b}h^{B\m\n\g} \6^2_{\m\n}h^C_{\a\b\g}
 - \frac{5}{4}\, h_a \6^{2\a\b}h^B_\m \6^{2\h\m}h^C_{\a\b\g}
 + \frac{5}{4}\, h_a \6^{2\a\b}h^B_{\a\m\g} \6^{2\m\n}h^{C\ \g}_{\b\n}
\nonumber \\
&&+\; \frac{1}{2}\, h_a^{\a\b} \6^2_{\a\b}h^{B\g} \Box h^C_\g
 - \frac{1}{4}\, h_a^{\a\b} \6^2_{\a\b}h^{B\m\n\r} \Box h^C_{\m\n\r}
 - \frac{3}{4}\, h_a^{\a\b} \6^2_{\a\b}h^{B\m\n\r} \6^2_{\m\n}h^C_\r
\nonumber \\
&&-\; \frac{1}{2}\, h_a^{\a\b} \6^2_{\a\b}h_\r^B \6^2_{\m\r}h^{C\m\n\r}
 - \frac{1}{8}\, h_a^{\a\b} \6^2_{\a\b}h^{B\m} \6^2_{\m\n}h^{C\n}
 + \frac{3}{4}\, h_a^{\a\b} \6^2_{\a\b}h^{B\r\s\m} \6^2_{\m\n}h^{C\ \n}_{\r\s}
\nonumber \\
&&+\; \frac{1}{2}\, h_a^{\a\b} \Box h^B_{\a\b\g} \Box h^{C\g}
 - \frac{1}{2}\, h_a^{\a\b} \Box h^B_{\a\b\g} \6^2_{\m\n} h^{C\g\m\n}
 - \frac{3}{4}\, h_a^{\a\b} \6^2_{\m\n} h^B_{\a\b\g} \Box h^{C\g\m\n}
\nonumber \\
&&-\; \frac{1}{2}\, h_a^{\a\b} \6^2_{\m\n} h^{B}_{\a\b\g} \6^{2\m\n} h^{C\g}
 + h_a^{\a\b} \6^2_{\m\n} h^{B}_{\a\b\g} \6^{2\m}_\r h^{C\g\n\r}
 + \frac{3}{4}\, h_a^{\a\b} \6^{2\m\n} h^{B}_{\a\b\m} \6^2_{\n\t}h^{C\t}
\nonumber \\
&&+\; \frac{3}{4}\, h_a^{\a\b} \6^{2\m\n} h^B_{\a\b\m} \Box h^C_\n
 - \frac{3}{4}\, h_a^{\a\b} \6^{2\m\n} h^B_{\a\b\m} \6^{2\r\s} h^C_{\n\r\s}
 - \frac{7}{4}\, h_a^{\a\b} \6^2_{\n\t} h^B_{\a\b\m} \6^{2\m\n} h^{C\t}
\nonumber \\
&&-\; \frac{1}{8} h_a^{\a\b} \Box h^B_{\a\b\m} \6^{2\m\n} h^C_\n
 + \frac{5}{4}\, h_a^{\a\b} \6^{2\r\s} h^B_{\a\b\m} \6^{2\m\n} h^C_{\n\r\s}
 - \frac{5}{4}\, h_a^{\a\b} \6^2_{\m\a} h^B_\b \Box h^{C\m}
\nonumber \\
&&+\; \frac{5}{4}\, h_a^{\a\b} \6^2_{\m\a} h^B_{\b\n\r} \6^{2\n\r} h^{C\m}
 + \frac{3}{4}\, h_a^{\a\b} \6^2_{\m\a} h^B_{\b\n\r} \Box h^{C\m\n\r}
 + \frac{5}{4}\, h_a^{\a\b} \6^2_{\m\a} h^B_\b \6^2_{\n\r} h^{C\m\n\r}
\nonumber \\
&&-\; \frac{5}{2}\, h_a^{\a\b} \6^2_{\m\a} h^B_{\b\n\r}\6_\t^{2\r}h^{C\n\r\t}
 + \frac{1}{4}\, h_a^{\a\b} \6^2_{\m\a} h^{B\m\n}_\b \6^2_{\n\r} h^{C\r}
 - h_a^{\a\b} \6^2_{\m\a} h^{B\m\n}_\b \Box h_\n^C
\nonumber \\
&&+\; h_a^{\a\b} \6^2_{\m\a} h^{B\m\n}_\b \6^{2\r\s} h^C_{\n\r\s}
 + h_a^{\a\b} \6^2_{\m\a} h^B_{\b\n\r} \6^{2\m\n} h^{C\r}
 - \frac{11}{8}\, h_a^{\a\b} \6^2_{\m\a} h^B_\b \6^{2\m\n} h^C_\n
\nonumber \\
&&+\; \frac{1}{4}\, h_a^{\a\b} \6^2_{\m\a} h^{B\r\s}_\b \6^{2\m\n} h^C_{\n\r\s}
 + \frac{5}{8}\, h_a^{\a\b} \6^2_{\m\a} h^{B\m} \Box h^C_\b
 - \frac{1}{4}\, h_a^{\a\b} \6^2_{\m\a} h^{B\m\n\r} \6^2_{\n\r} h^C_{\b}
\nonumber \\
&&+\; \frac{1}{2}\, h_a^{\a\b} \6^2_{\m\a} h^{B\m\n\r} \Box h^C_{\b\n\r}
 - \frac{3}{4}\, h_a^{\a\b} \6^2_{\m\a} h^{B\m} \6^{2\n\r} h^C_{\b\n\r}
 - h_a^{\a\b} \6^2_{\m\a} h^{B\m\n\r} \6^2_{\n\t} h^{C\ \t}_{\b\r}
\nonumber \\
&&+\; h_a^{\a\b} \6^2_{\m\a} h^{B\t}\6_{\n\t}^2 h_\b^{C\m\n}
 - h_a^{\a\b} \6^2_{\m\a} h^B_\n \Box h_\b^{C\m\n}
 + h_a^{\a\b} \6^2_{\m\a} h^{B\t} \6^{2\m\n} h^C_{\b\n\t}
\nonumber \\
&&+\; \frac{1}{2}\, h_a^{\a\b} \6^2_{\m\a} h^B_\n \6^{2\m\n} h^C_\b
 - \frac{1}{2}\; h_a^{\a\b} \6^2_{\a\m} h^{B\n} \6^{2\m}_\b h^C_\n
 - \frac{1}{4}\; h_a^{\a\b} \6^2_{\a\m} h^{B\n\r\s} \6^{2\m}_\b h^C_{\n\r\s}
\nonumber \\
&&+\; \frac{1}{4}\, h_a^{\a\b} \6^2_{\a\m} h^{B\m} \6^2_{\b\n} h^{C\n}
 - \frac{1}{4}\, h_a^{\a\b} \6^2_{\a\m} h^{B\m\r\s} \6^{2\n}_\b h^C_{\n\r\s}
 - \frac{1}{2}\, h_a^{\a\b} \6^2_{\a\m} h^{B\n} \6^2_{\b\n} h^{C\m}
\nonumber \\
&&+\; \frac{1}{4}\, h_a^{\a\b} \6^2_{\a\m} h^{B\n\r\s} \6^2_{\b\n} h^{C\ \m}_{\r\s}
 + h_a^{\a\b} \6^2_{\a\m} h^{B\m\n\r} \6^2_{\b\n} h_\r^C
 + \frac{1}{4} h^{\a\b}_a \Box h^B_\a \Box h^C_\b
\nonumber \\
&&-\; \frac{1}{4}\, h^{\a\b}_a \Box h^{B\m\n}_\a \Box h^C_{\b\m\n}
 + \frac{1}{4}\,h^{\a\b}_a \Box h^B_{\a\m\n} \6^{2\m\n} h^C_\b
 - \frac{3}{4}\,h^{\a\b}_a \Box h^B_\a \6^{2\m\n} h^C_{\b\m\n}
\nonumber \\
&&+\; h^{\a\b}_a \Box h^{B\ \r}_{\a\m} \6^2_{\n\r} h^{C\m\n}_\b
 + \frac{1}{2}\, h^{\a\b}_a \6^{2\m\n} h^B_{\a\m\n} \6^{2\r\s} h^C_{\b\r\s}
 - \frac{1}{2}\, h^{\a\b}_a \6^{2\m\n} h^B_{\a\r\s} \6^{2\r\s} h^C_{\b\m\n}
\nonumber \\
&&-\; \frac{1}{2}\, h^{\a\b}_a \6^{2\m\n} h^B_{\a\m\r} \6^{2\r\s} h^C_{\b\n\s}
 + \frac{1}{2}\, h^{\a\b}_a \6^{2\m\n} h^B_\a \6^2_{\m\n} h^C_\b
 - \frac{1}{2}\, h^{\a\b}_a \6^{2\m\n} h^{B\r\s}_\a \6^2_{\m\n} h^C_{\b\r\s}
\nonumber \\
&&-\; \frac{1}{2}\, h^{\a\b}_a \6^{2\m\r} h^B_{\a\r\s} \6^{2\s}_\m h^C_\b
 - \frac{1}{2}\, h^{\a\b}_a \6^{2\m\r} h^B_{\a\r\t} \6^2_{\m\s} h_\b^{C\s\t}
 + h^{\a\b}_a \6^{2\m\r} h_\a^{B\s\t} \6^2_{\m\s} h^C_{\b\r\t}\,.
\nonumber
\end{eqnarray}

\end{document}